\documentclass[review]{elsarticle}

\usepackage{lineno,hyperref}
\usepackage{cleveref} 

\usepackage{graphicx, rotating}
\usepackage{placeins}  
\usepackage{subcaption} 
\usepackage{boxedminipage} 
\usepackage{multirow} 
\usepackage{setspace}  

\usepackage{supertabular}  
\usepackage[english]{babel}  

\usepackage{color,soul} 

\journal{Journal of \LaTeX\ Templates}

\biboptions{longnamesfirst,semicolon,sort&compress}
\begin{document}

\begin{frontmatter}

\title{Experimental study of concrete activation compared to MCNP simulations for safety of neutron sources}


%
\author[mymainaddress,mysecondaryaddress]{D.~Hajd\'{u}\corref{mycorrespondingauthor}}
\cortext[mycorrespondingauthor]{Corresponding author}
\ead{hajdu.david@energia.mta.hu}
\author[mymainaddress,mytertiaryaddress,myquaternaryaddress]{E.~Dian}
\author[mymainaddress]{K.~Gm\'{e}ling}
\author[mytertiaryaddress,myquinternaryaddress]{E.~Klinkby}
\author[mytertiaryaddress] {C.~P.~Cooper-Jensen}
\author[mymainaddress,mysenaryaddress]{J.~Os\'{a}n}
\author[mymainaddress,myquaternaryaddress]{P.~Zagyvai}


\address[mymainaddress]{Hungarian Academy of Sciences, Centre for Energy Research, 29-33.~Konkoly-Thege M., 1525~Budapest, Hungary}
\address[mysecondaryaddress]{University of Pannonia, Institute of Radiochemistry and Radioecology, 10.~Egyetem, 8200~Veszpr\'{e}m, Hungary}
\address[mytertiaryaddress] {European Spallation Source ESS ERIC, P.O Box 176, 221~00~Lund, Sweden}
\address[myquaternaryaddress]{Budapest University of Technology and Economics, Institute of Nuclear Techniques, 9.~M\H uegyetem rakpart, 1111~Budapest, Hungary}
\address[myquinternaryaddress]{Technical University of Denmark, DTU Physics Ris\o Campus, 399~Frederiksborgvej, 4000~Roskilde, Denmark}
\address[mysenaryaddress]{International Atomic Energy Agency (IAEA), Nuclear Science and Instrumentation Laboratory, IAEA Laboratories, 2444~Seibersdorf, Austria}

\begin{abstract}
The neutron activation of shielding materials and the generated decay gamma radiation are well-known issues in terms of occupational exposure. Though the trace elements of shielding concretes can be dominant sources of the produced activity in such cases, their concentrations are often missing from the input data of shielding-related Monte Carlo simulations. For this reason, three concrete types were studied, that were considered in the European Spallation Source (ESS) ERIC. Their composition –- including the trace elements -- were determined via XRF, PGAA and NAA techniques. Realistic input data were developed for these materials, containing the parent elements of all the dominant radioisotopes, and were validated against measured data of neutron irradiation experiments. 
\end{abstract}



\end{frontmatter}


\section{Introduction}

The European Spallation Source (ESS) ERIC~\cite{ESSERIC} -- which is now under construction -- will be the brightest neutron source of the world~\cite{Peggs2013}. Due to the high neutron fluxes with wide variety of energies in the bunker, the different neutron guides and the experimental caves, an extensive and detailed safety planning is essential comprising all components of the facility that will be exposed to radiation. 
The occurring radiation fields will have several different components, e.g.\ primary and scattered thermal and fast neutron radiation~\cite{Dijulio2016, Cherkashyna2015a, Cherkashyna2015, Cherkashyna2014} and neutron induced prompt- and decay gamma radiation. Most of these components have already been studied in terms of radiation protection~\citep{Mantulnikovs2016} and radioactive waste production~\cite{Kokai2018}. Due to the upcoming hot commissioning of the ESS, the scope of the current study is the expected occupational exposure of maintenance personnel.
In typical maintenance scenarios the main source of radiation is the remnant activity from the neutron activation of various instrument components, e.g.\ choppers, detectors, neutron guides and shielding. The activation of many of these components have already been studied~\cite{Kokai2018, Kokai2017, Dian2017}, thus the focus of the current work is set on the activation of shielding materials, specifically on the widely used concrete.
As in conventional radiation protection the primary focus is set on the shielding properties of materials, most of the concrete compositions that are currently available for simulation purposes are limited to the major components of the concrete~\cite{McConnJr2011}. This indeed determines the shielding properties, but certain trace elements are expected to be the dominant sources of neutron induced radiation due to their particularly high cross sections for neutron activation~\cite{Tesse2018, Yarar1996}. Therefore, a detailed knowledge of the amount and effects of trace elements in shielding materials is necessary for a responsible safety planning in such cases.

In this study three concrete types were examined, that are relevant to the ESS: \textit{i)} a general concrete produced by Skanska company~\cite{SkanskaSE} -- referred to as `Skanska concrete' --, which is used in the construction of the ESS; \textit{ii)} the PE-B4C-concrete~\cite{DiJulio2017}, which has enhanced neutron shielding properties~\cite{Park2014a, DiJulio2018, Cai2018, El-Sarraf2013, Abrefah2011} due to its polyethylene and boron-carbide additives, and which was developed particularly for the ESS. And finally \textit{iii)} the ordinary concrete from which PE-B4C-concretete was developed -- called as `Reference concrete' in the followings. 
The activation characteristics of these concretes were determined via neutron irradiation experiments in the Budapest Research Reactor (BRR)~\cite{BNC}, and these measured data were used to validate the detailed input data for realistic activation calculations. For this purpose, the initial elemental compositions of the concrete samples were measured with different analytical techniques. Prompt Gamma Activation Analysis (PGAA)~\cite{Szentmiklosi2010} was used along with Neutron Activation Analysis (NAA)~\cite{Szentmiklosi2016} as the most suitable option for the identification of highly neutron absorbing elements. Due to its accessibility and rather different range of component sensitivity X-ray fluorescence (XRF) measurements were also performed.
The measured compositions then were used as inputs in MCNPX~\cite{MCNPX} and Cinder1.05~\cite{Cinder90} simulations. With the first set of simulations, the irradiations in the BRR were reproduced and the generated radioisotopes were compared against the measured results. Then, in a realistic ESS bunker maintenance scenario, dose consequences of the shielding concrete activation were compared applying the nominal composition (given by the manufacturer) and our measurement-based composition, highlighting the importance of trace elements.

\section{Methodology}

Three types of concrete samples: PE-B4C-concrete, its Reference concrete and a Skanska concrete were examined in the current study. The samples were received as grist. Elemental compositions of the samples were measured with NAA, PGAA and XRF analytical techniques. During the NAA experiments, activation properties of the samples were also assessed. Based on these experiments, simplified but realistic concrete compositions were determined and validated for activation simulation purposes (presented in `material card' format for MCNP simulations). Finally, an ESS bunker case study was performed to demonstrate the applicability and importance of the improved material cards in safety planning. The main focus of the current study is on the decay gamma emission of the irradiated shielding concretes, and its potential dose consequences in terms of occupational exposure. For this reason, -- considering the high self-attenuation of concrete for $\alpha$- and $\beta$-radiation, -- in the followings only the contribution of gamma-emitting activation products are discussed.

\subsection{NAA measurements}\label{M_mNAA}

For the NAA analysis $\sim$0.1~g samples  of each concrete were filled to high-purity quartz ampoules, measured by a Mettler-Toledo XPE 26 microbalance having $<$0.7~$\mu$g reproducibility. Two sets of samples were prepared and encapsulated in 10~cm long aluminum tubes, and irradiated in the water-filled, vertical irradiation channels of the 
BRR~\cite{BNC}. The first set was irradiated in a rotated, well-thermalized irradiation channel (called `Thermalized channel' in the followings) located in the beryllium reflector around the reactor core. The second set was placed into an irradiation channel in the reactor core with higher fast/thermal neutron flux ratio (referred to in the followings as `Fast channel'). The irradiation time was 2~hours for both cases. Neutron flux parameters during the irradiation were measured with Bare Triple-Monitor method~\cite{Corte1979a} and are presented in Table~\ref{tab:irrChflux}. The boundaries of epithermal neutrons are 0.5 and 100~eV.

\begin{table}[htbp]
  \centering
  \caption{Measured neutron fluxes in the vertical irradiation channels of BRR~\cite{BNC}. Energy boundaries of epithermal flux: 0.5 and 100~eV. (The uncertainties are 15-20\%.) }
  \label{tab:irrChflux}
  \resizebox{0.75\textwidth}{!}{
    \begin{tabular}{lcc}
      \hline
                                    & Thermalized channel & Fast channel \\
      \hline
      Thermal flux~[cm$^{2}$/s]     & 2.00E+13           & 5.00E+13      \\
      Epithermal flux~[cm$^{2}$/s]  & 4.30E+11           & 3.80E+12      \\
      Fast flux~[cm$^{2}$/s]        & 1.30E+12           & 4.70E+13      \\
      \hline
  \end{tabular}}
\end{table}

The activities of the irradiated samples were determined by measuring their gamma response with a Canberra HPGe detector (p-type detector, 36\% relative efficiency, 1.75~keV/1332.5~keV resolution connected to Ortec 502 MCA-2) in a low background measurement chamber. The first measurements were performed after 4~days of cooling due to the high initial activity of the samples. Samples were measured 5~times in a 3-week long period, to determine their decay characteristics. The measurement time increased from 10~minutes to 2~hours in this period. Gamma spectra were evaluated with Hypermet PC~\cite{Szentmiklosi2016} and KayZero for Windows 3.06 programs~\cite{DeCorte2001} resulting  elemental compositions.

The measured gamma spectra were not only used for elemental composition analysis, but also to characterise the activation properties of the concrete samples, especially comparing the activation and decay profiles of the PE-B4C-concrete with that of the Reference concrete. 
In addition, in order to facilitate the comparison of the measured results with simulated activity concentrations -- as Cinder1.05 output tends to give an overflowed, hardly searchable inventory-- so called `key isotopes' were identified in the measured spectra, giving the majority 
of the total activity in all measured and extrapolated activities. 
All parent elements of these key isotopes were also identified, and were in the focus of the following composition measurements.

\subsection{PGAA measurements}

Prompt gamma activation analysis was performed on all three types of concrete samples at the PGAA Experimental Station of the BRR~\cite{Szentmiklosi2010}. In the experiment 6~g samples of each concrete was irradiated with a 20~$\times$~20~mm collimated cold neutron beam~\cite{Rosta1997}. The emitted gamma-rays were detected with HPGe detector surrounded by a bismuth germanate (BGO) scintillator detector. Signals were processed and evaluated with a Canberra AIM 556A multichannel analyser and Hypermet-PC gamma software~\cite{Fazekas1997}. Element identification was performed with the ProSpeRo program, utilising the prompt gamma analysis libraries from Molnar et al.~\cite{Molnar2004a}. In this PGAA station 16~elements could be quantified: H, C, Na, Al, Si, S, K, Ca, Ti, Mn, Fe, B, Cl, V, Sm and Gd.

\subsection{XRF measurements}



In addition to the above described methods, as a relatively fast and low-cost technique, X-ray fluorescence (XRF)~\cite{West2016} spectroscopy was also carried out on the concrete samples for elemental analysis  in the Nuclear Science and Instrumentation Laboratory of the IAEA~\cite{IAEASeibersdorf}.

To increase the homogeneity of the grist samples, they were ground with a ball grinder using tungsten carbide (WC) balls and mortars. The material used for grinding was considered during evaluation of elemental concentrations. After the addition of 0.25~g wax to each sample, they 
were pressed to 2.5~g pellets. Three samples were prepared from all of the concretes, and all samples were measured with an Epsilon 5 (PANalytical, The Netherlands) triaxial polarising energy-dispersive XRF (EDXRF) device~\cite{Epsilon5}. The instrument is equipped with a 100-kV X-ray tube, a series of secondary targets and a HPGe detector with high efficiency for hard X-rays. Each sample was measured under ten different conditions using secondary targets suitable for different groups of elements. Concentrations were calculated using the built-in software based on all ten spectra for each sample. Calibration curves generated from measurements of several geological certified reference materials were used for the concrete samples. Using several secondary targets is advantageous since it allows a full elemental analysis of the samples for elements from Na to U, considering major elements as present in the common oxide forms~\cite{Cevik2013}. The condition using a 100-keV excitation and a Barkla polarising secondary target makes it possible to quantify lanthanide elements based on their K lines. EDXRF in general has a high dynamic range covering major to trace elements.  

The composition of the so-called dark matrix was considered from the nominal composition of the samples, given by the manufacturer. Preliminary results of PE-B4C and Reference concretes have already been published elsewhere~\cite{Dian2018}.

Because of the strong spectral overlapping between Ca-K$\beta$ and Sc-K$\alpha$ as well as Fe-K$\beta$ and Co-K$\alpha$, elements Sc and Co were determined using wavelength-dispersive XRF (WDXRF). An ARL Optim'X spectrometer (Thermo Scientific~\cite{WDXRF}) was applied for this task.

\subsection{Simulations of concrete activation measurements}

\begin{figure}[ht!]
\centering
  \includegraphics[width=0.15\textwidth]{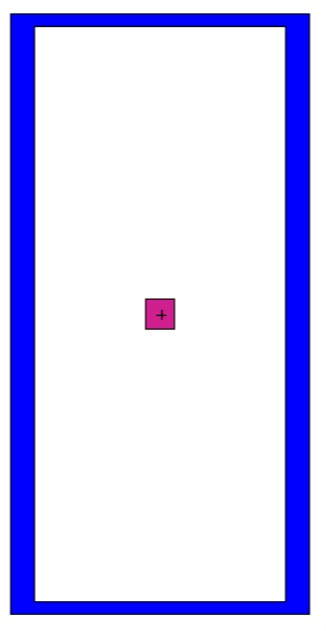}
  \caption{Irradiation geometry in MCNPX simulations with blue aluminium sample holder (10~cm long tube with 2.1~cm inner radius and 0.4~cm wall thickness), purple sample ($\rm 0.5 \times 0.5 \times 0.5~cm$ cube) and white void in between. \label{irrgeom} }  
\end{figure}

MCNPX neutron activation simulations were performed in order to study the impact of the concrete composition on the produced activity. For this purpose, the neutron irradiation experiments (see Section~\ref{M_mNAA}) were reproduced via simulations. The constructed geometry consisted of the sample and the aluminium sample holder (see Figure~\ref{irrgeom}) thus placed in a $\rm 20 \times 20 \times 20~cm$ cube, representing a homogeneous volumetric neutron source. 

\begin{figure}[ht!]
  \centering
  \includegraphics[width=0.75\textwidth]{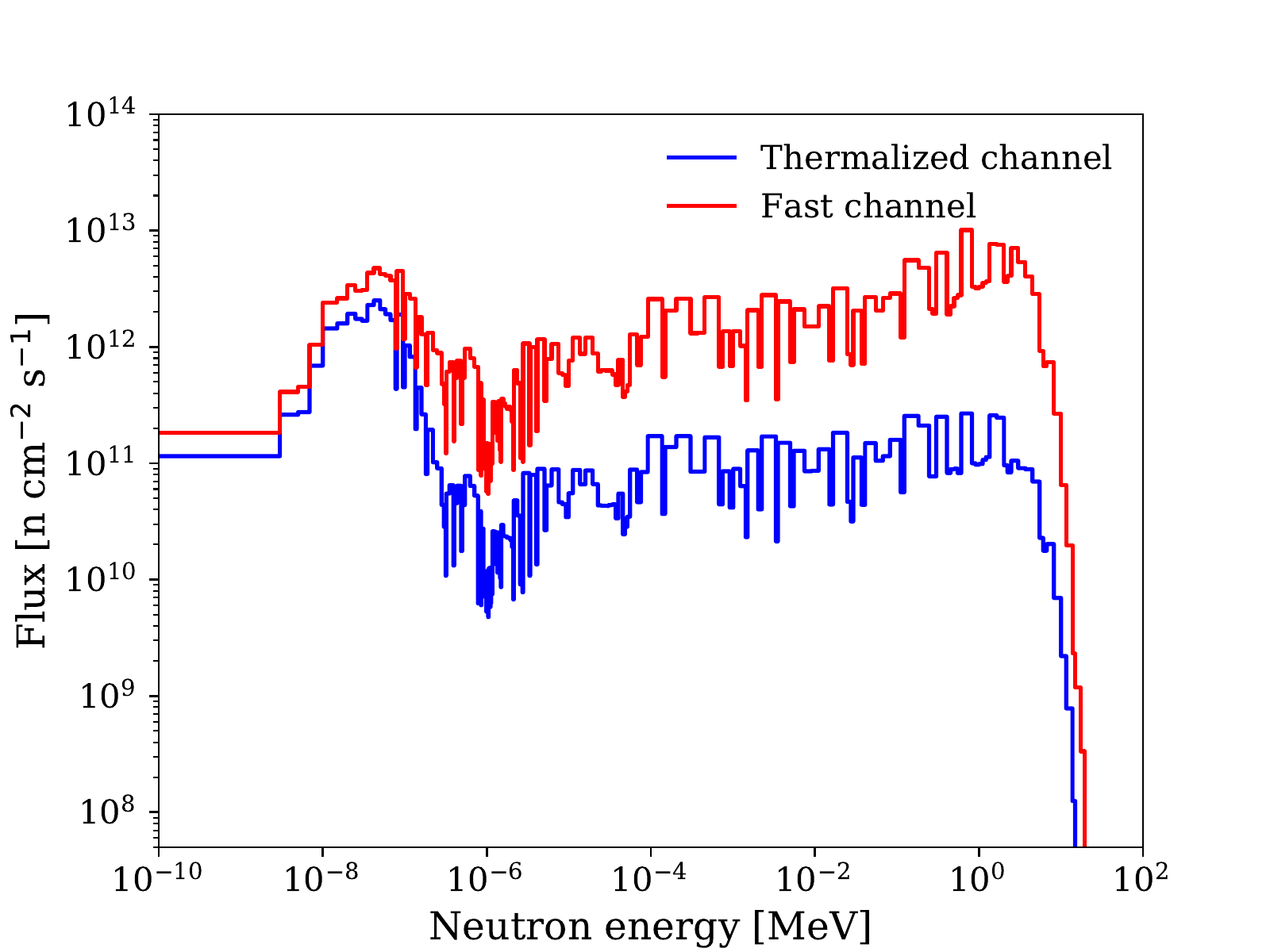}
  \caption{Typical neutron spectra in the `Thermalized' and `Fast' vertical irradiation channels of the BRR~\cite{BNC}. \label{irrflux} }  
\end{figure}

The experiments in both the Fast and the Thermalized channels were reproduced in simulations,  with their typical neutron spectra which are presented in Figure~\ref{irrflux}. The default spectral distributions of the channels were adjusted to the current measurements by scaling them with the measured flux parameters presented in Table~\ref{tab:irrChflux}.

Simulations were performed with different initial compositions (i.e.\ `material cards') for all three concretes: \textit{i)} the nominal composition, having similar level of details as the currently available MCNP concrete material cards, e.g.\ the Los Alamos (MCNP) material card~\cite{McConnJr2011}, \textit{ii)} the composition obtained from EDXRF and \textit{iii)} the composition based on the NAA and PGAA measurements. The results of the PGAA and the NAA methods were considered in combination, as they are practically complementary methods. The measurement-based compositions involved all elements indicated in the nominal composition, extended with the parent elements of the key radioisotopes. Nominal values were applied for non-detectable bulk elements. 

Irradiation conditions were reproduced using  MCNPX~\cite{MCNPX} and  Cinder1.05~\cite{Cinder90} codes. Activity concentrations were calculated for the end of the respective irradiation time (t$_{0}$) and also for the times of the activity measurements. Considering radioactive waste production, the activity concentrations of the samples were determined for 1~year and 5~years of cooling as well. The simulated activity concentrations were compared to the measured ones for the whole described time period, where the measured activities were extrapolated to  t$_{0}$ and 1~y and 5~y cooling times. While the focus of the current study is on the gamma-emitting isotopes, Cinder provides a detailed but indiscriminate inventory of all activation products. For the fast and easy comparison of the measured and simulated data, the total simulated activity concentrations were approximated as the sum of the activity concentrations of the key isotopes, described in Section~\ref{M_mNAA}.

On the basis of the measured compositions and the activity simulations, combined input compositions  were constructed for the PE-B4C and the Reference concrete that sufficiently reproduced the measured activities, and therefore recommended for further activation simulations for e.g.\ radiation safety planning. These compositions consist of two kinds of elements: major elements indicated in the nominal compositions (H, C, Na, Al, Si, S, K, Ca, Fe, B, Mg, Ti, P, Cl) and the parent elements of the key isotopes. For each element a measured value was preferred if it was available, otherwise the nominal one was used. As (in this unique case) the EDXRF analyses were performed with the most number of samples, the EDXRF produced values were favoured as default against other measured results, considering the better statistics. However, NAA produced values were considered representative for elements in very low amount ($10^{-4}$~w\%), as in this case NAA provided lower detection limit than EDXRF, giving presumably more precise values. In addition, when the different methods provided conflicting results -- mostly due to the general inhomogeneity of the samples --, the more conservative values were selected. Finally, as the oxygen content varies significantly in the concrete, it was chosen to assume that the mass fraction unaccounted for by the above consisted of oxygen. The material compositions created this way, i.e.\ the \textit{iv)} recommended compositions were validated via the above described MCNPX simulations.

\subsection{Dosimetrical simulations in the ESS bunker}

\begin{figure}[ht!]
  \centering
  \includegraphics[width=0.65\textwidth]{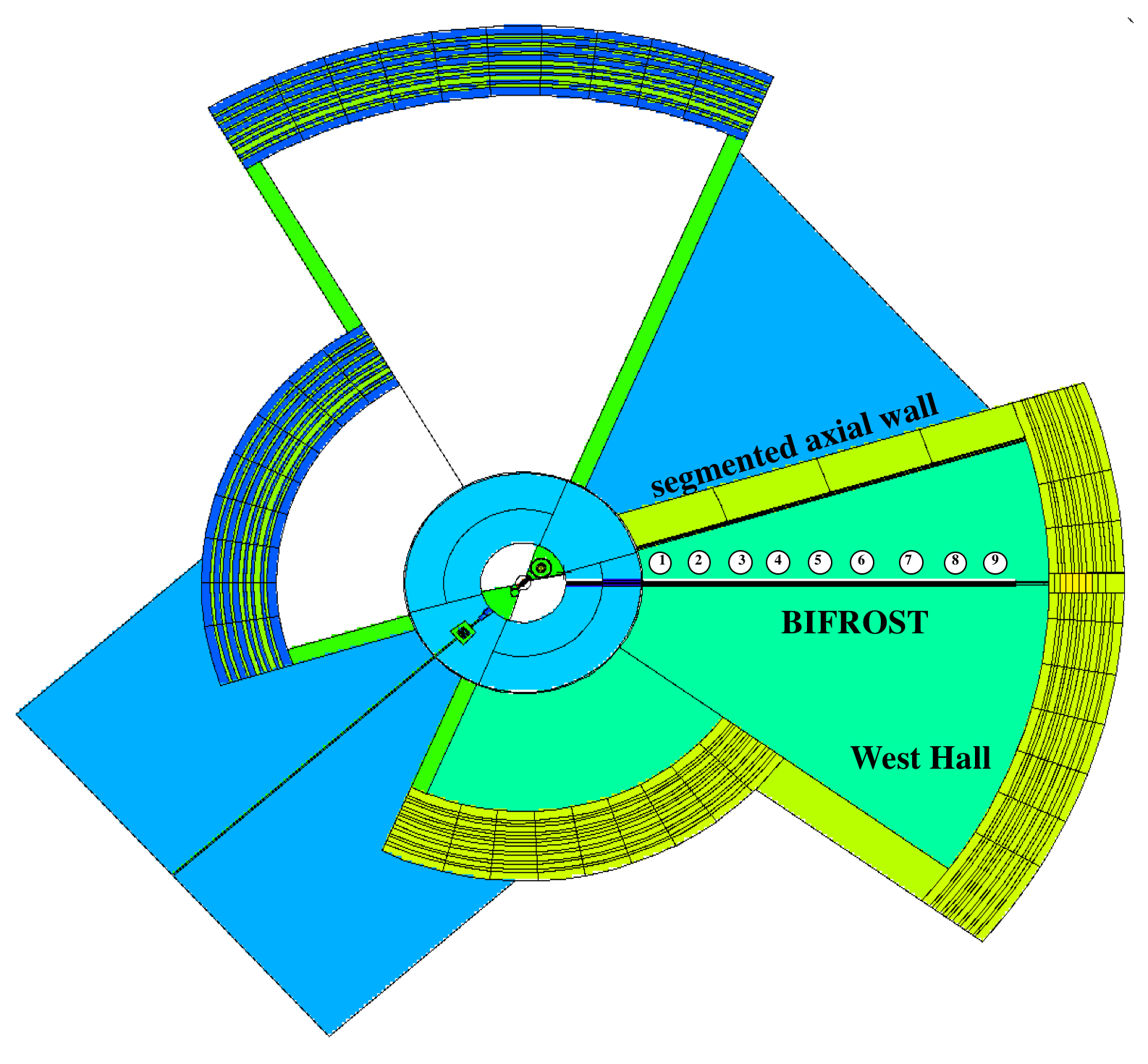}
  \caption{Overview of the MCNP model of the bunker complex. 1-9: 50~cm radius spheres along the neutron guide of the BIFROST instrument, marked areas for numerical comparison of decay gamma dose rates. Source of the radiation was the activated segmented axial wall made of Reference concrete. \label{bunker_geom} }  
\end{figure}

A set of simulations was also performed in order to evaluate the impact of realistic concrete composition from dosimetric aspects. Decay gamma dose rate coming from the irradiated concrete shielding was determined with nominal and measurement-based recommended compositions. The West Hall of the ESS's bunker was selected as a test case, as described in~\cite{Bunker2018}. For this purpose, a detailed MCNP model of the bunker complex was developed, including the penetration through the monolith and guide segment corresponding to the BIFROST instrument, as shown in Figure~\ref{bunker_geom}. The ESS bunker is planned to be built from two types of concrete: a heavy concrete for the radial walls and the roof, and a normal concrete for the axial wall. For the similarity with the real case of the ESS design, the Reference concrete was applied in the model in the axial walls. The neutron activation process and 
generated decay gamma dose rates were modelled for only this single wall, eliminating the additional decay gamma background originating from the heavy concrete walls.

\FloatBarrier

The simulations were performed with the combination of MCNPX and Cinder1.05 code in a realistic operational scenario of the ESS: 10~years of operation on average power ($\rm 0.616 \times 5$~MW) and 6~month operation in full power (5~MW) followed by three days of cooling, representing e.g.\ a maintenance period. For the MCNP transport the source was constituted by 2~GeV protons impinging on the tungsten target. In turn, neutrons of different energies and directions were generated in the collisions. Then the neutron flux and isotope production rates were determined for the axial wall, divided into 36~segments considering the effect of self-shielding. Cinder1.05 code was used to determine the production of radioisotopes, providing the corresponding Gamma source description with the usage of Gamma Script. This source was applied again in a second set of MCNP simulation, to determine the decay gamma dose rates -- originating from the axial wall made of the Reference concrete  -- for the whole West Hall of the ESS bunker. In addition, the dose rates were recorded in 9~spherical areas -- shown is Figure~\ref{bunker_geom} -- for numerical comparison.

\section{Results and Discussion}

\subsection{Concrete activity measurements}

\begin{figure}[ht!]
  \centering
  
  \begin{subfigure}[b]{0.7\textwidth}
    \includegraphics[width=\textwidth]{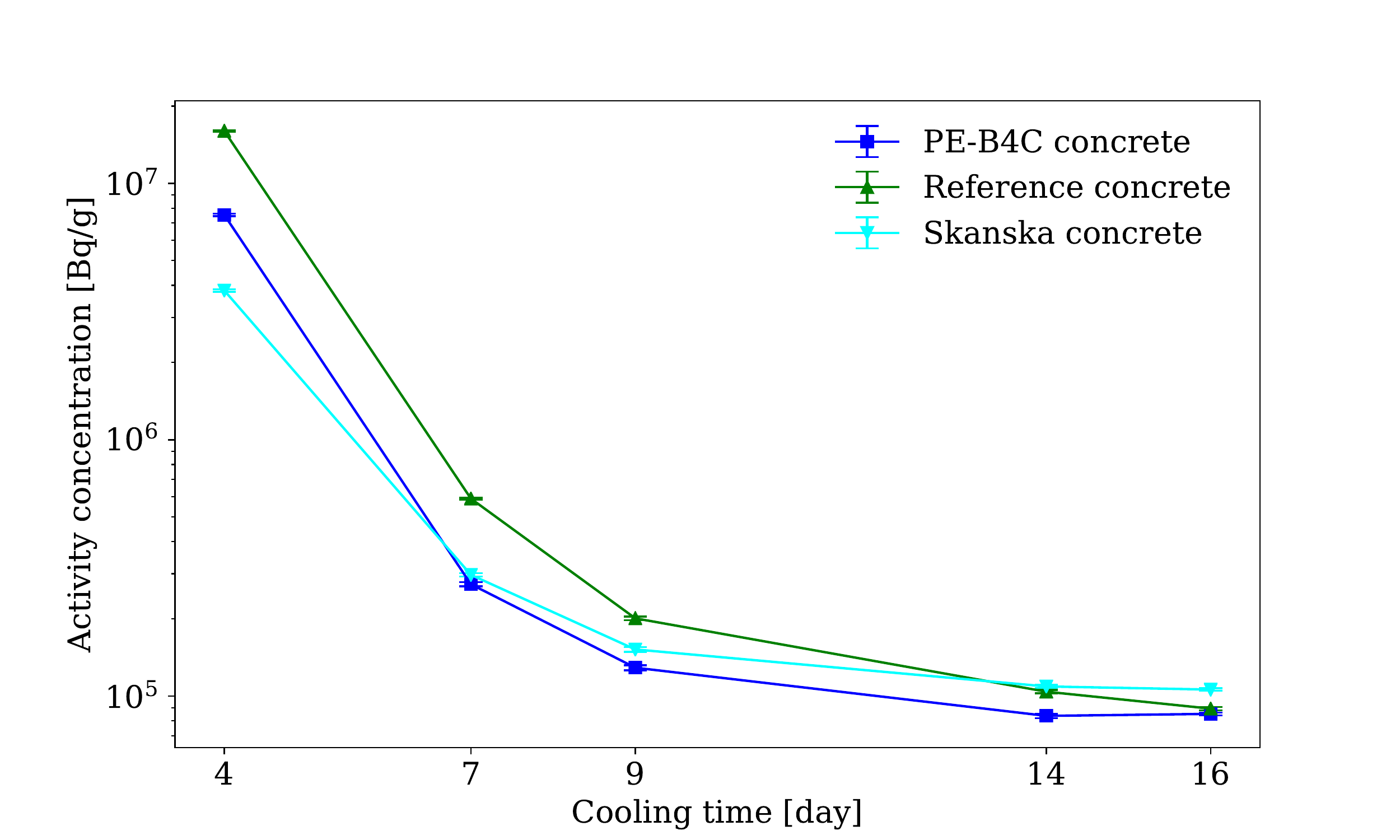}
    \caption{Thermalized channel\label{m_Ac_th}}
  \end{subfigure}
  ~
  \begin{subfigure}[b]{0.7\textwidth}
    \includegraphics[width=\textwidth]{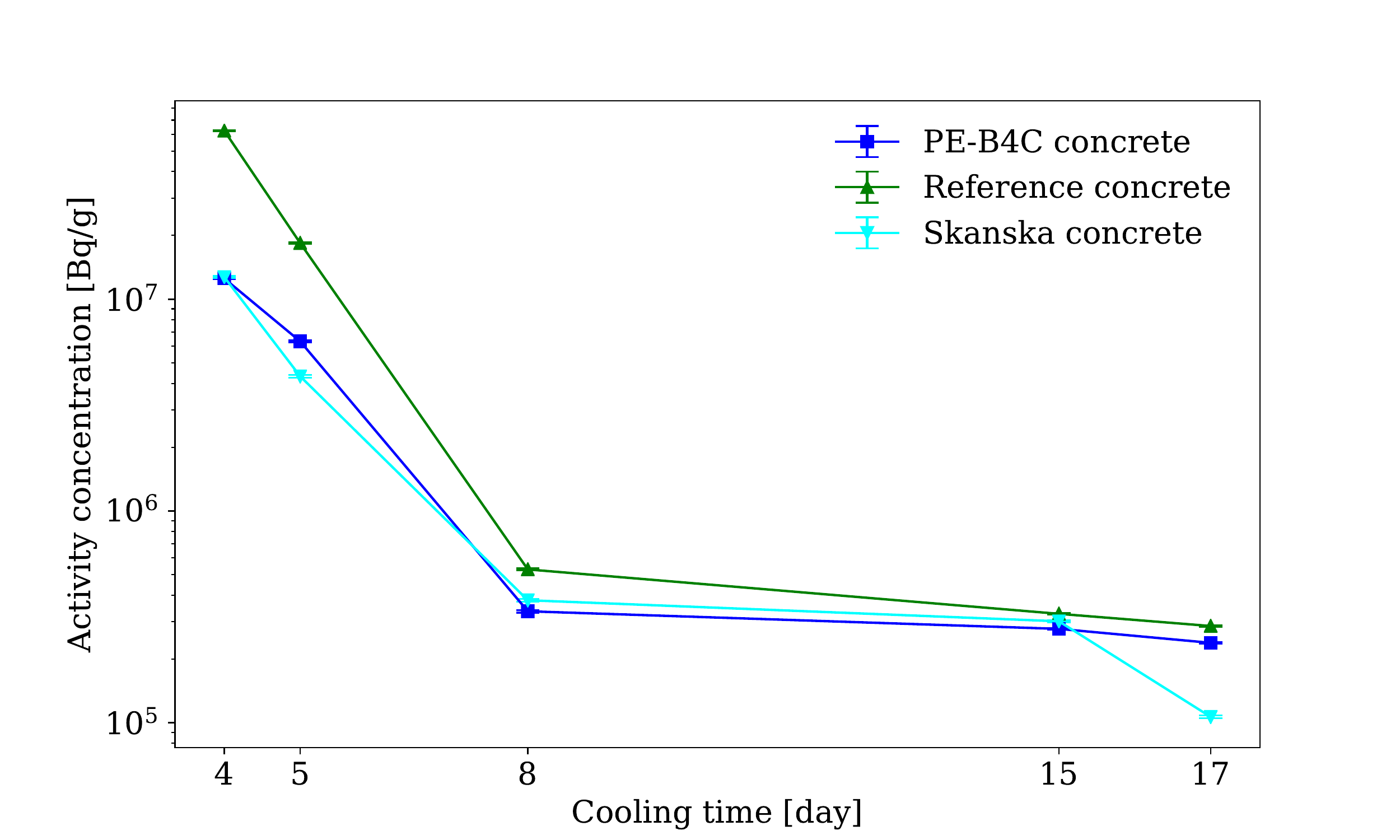}
    \caption{Fast channel\label{m_Ac_f}}
  \end{subfigure}

  \caption{Measured decay profiles of the three concrete samples after irradiation in the Thermalized~(\ref{m_Ac_th}) and Fast channels~(\ref{m_Ac_f}). The statistical uncertainties are too small to be discernible. The measured points are connected for better visibility.\label{m_AcComp} }
\end{figure}

The activities of the three studied concrete samples were determined after irradiation in the Fast and Thermalized vertical irradiation channels of the BRR, with the main goal of comparing their total activity in a 2-week long follow-up period. 
The total activities of the samples were calculated as the sum of the activities of all the determined gamma-emitting radioisotopes in their respective measured spectra. The measured decay curves of the three concretes are presented in Figure~\ref{m_AcComp} for both irradiation channels.

It is shown that, in accordance with the expectations, the measured total activity of the newly developed PE-B4C-concrete is consistently lower than that of the Reference concrete for the whole studied time period in both sets of measurements, and significantly lower during the first week after irradiation. PE-B4C-concrete produced $\sim$53-54\% and 66-80\% lower activity during the first week of cooling for the Thermalized and Fast channels, respectively. For this reason, PE-B4C-concrete is proven to be more advantageous in terms of neutron induced decay gamma emission, especially for a few days of cooling, i.e.\ in case of maintenance. Comparing the decay profiles of the PE-B4C- and Reference concrete with that of the Skanska concrete, it is also revealed that the reduction of the neutron-induced activity of the Skanska concrete, compared to the Reference concrete, is in the same range that of the PE-B4C-concrete: $\sim$50-76\% and 76-80\% during the first week of cooling for the Thermalized and Fast channels, respectively. This indicates that the remnant activity of a concrete can be significantly reduced with the application of polyethylene and boron-carbide, although, the variation of activity production due to the different concrete types can already be within the same range, highlighting the importance of the initial composition.

\begin{figure}[ht!]
  \centering
  \includegraphics[width=0.8\textwidth]{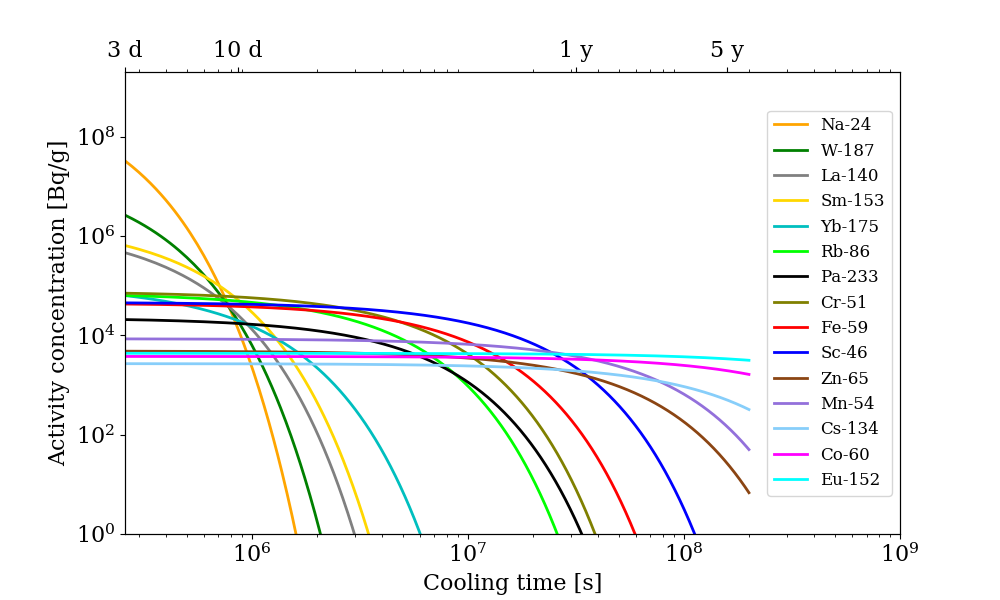}
  \caption{Decay curves of the 15~key radioisotopes of PE-B4C-concrete after irradiation in the Fast channel. Extrapolated from measured data. \label{15decay} }  
\end{figure}

In addition, in order to provide guidelines for simplified but realistic Monte Carlo activation simulations, and to enlighten the comparison of measured and simulated data, a set of 15~key gamma-emitting isotopes was determined, giving the majority (minimum 83\%) of the total activity for all the measured data in all samples and both irradiation channels. Both short- and long-lived isotopes are selected, giving a sufficient estimation of the total activity considering the potential occupational exposure after the end of irradiation, and for radioactive waste management after a typical 5-year cooling. The key radioisotopes and their typical activity concentrations in the studied time period -- extrapolated to 5~years of cooling  -- are presented in Figure~\ref{15decay}. Among the 15 isotopes, $^{24}$Na and $^{187}$W were found to be dominant in the first few days of cooling, while $^{60}$Co and $^{152}$Eu are giving the majority of the total activity in long-term, in accordance with the results of similar studies~\cite{Alhajali2016, Stefanov1989}. Activity simulations are expected to include these key isotopes as neutron activation products. 

For this purpose, potential parent elements of the radioisotopes shown in Figure~\ref{15decay} were identified. Most of them are produced via (n,$\gamma$) reaction, with a few exceptions. Beside (n,$\gamma$) reaction, $^{46}$Sc can also be a product of fast neutron (n,p) reaction of the 8\% natural $^{46}$Ti content of titanium. Similarly, $^{54}$Mn is produced as a fast neutron (n,p) reaction product of $^{54}$Fe, and $^{24}$Na also can be produced via a fast reaction from $^{27}$Al, in addition to the (n,$\gamma$) reaction of $^{23}$Na. The parent element of $^{233}$Pa is thorium, as it is the decay product of short-lived $^{233}$Th, produced by the (n,$\gamma$) activation of natural thorium. In total 14 parent elements ($^{54}$Fe and $^{54}$Mn are both originating from iron) were determined and recommended to be accounted for the initial concrete compositions. 

\FloatBarrier

\subsection{Composition measurements}

In order to provide detailed input data for Monte Carlo neutron activation simulations, the composition of the three concretes were determined via PGAA, NAA and EDXRF analytical measurements. The results of all samples are presented in \Cref{tab:compPE,tab:compRef,tab:compSka}. It is revealed that in accordance with the expectations, the combination of  PGAA and NAA measurements provided detailed and reliable data of the composition. PGAA mostly provided the bulk element content, while NAA the trace elements. It must be noted, that the NAA method may be less accurate compared to EDXRF in case of those parent elements which were activated to the same radioisotopes in different reaction routes (e.g. Na and Al). 

EDXRF measurements were also performed, providing input data both on bulk and trace elements beyond Na. Energy-dispersive X-ray fluorescence spectroscopy measurements show good agreement with the results of activation analysis for most of the elements. However, the parent elements of a few key isotopes were not measurable:  Sm and Yb were under detection limit of EDXRF, of~2~and~20~ppm, respectively. Furthermore, the two most important elements generating long-lived isotopes were not detected either: 
Eu concentration found to be under the detection limit of 1~ppm and Co was not measurable due to severe spectral overlapping with the high iron concentration of the samples. For this reason, although EDXRF is an easy and relatively cheap technique for the quality assurance of concretes, it may not be feasible for characterizing shielding materials which will be exposed to high neutron flux. The use of wavelength-dispersive XRF (WDXRF) overcome the difficulty with cobalt, scandium and europium due to its high spectral resolution.

\subsection{Monte Carlo modelling of sample activation}

\begin{figure}[ht!]
  \centering
  
  \begin{subfigure}[b]{0.48\textwidth}
    \includegraphics[width=\textwidth]{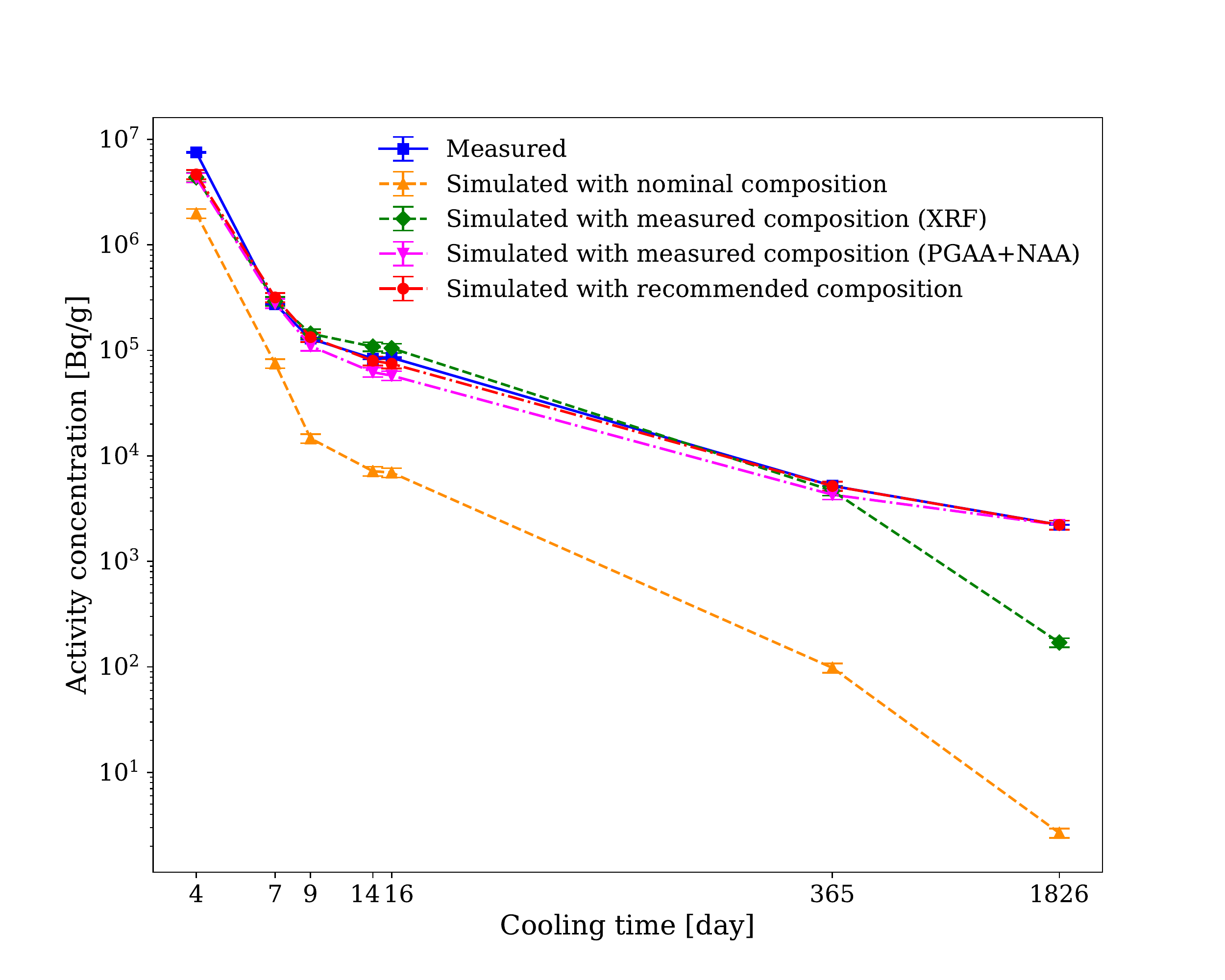}
    \caption{\label{matComp_th_PEB4C}}
  \end{subfigure}
  ~
  \begin{subfigure}[b]{0.48\textwidth}
    \includegraphics[width=\textwidth]{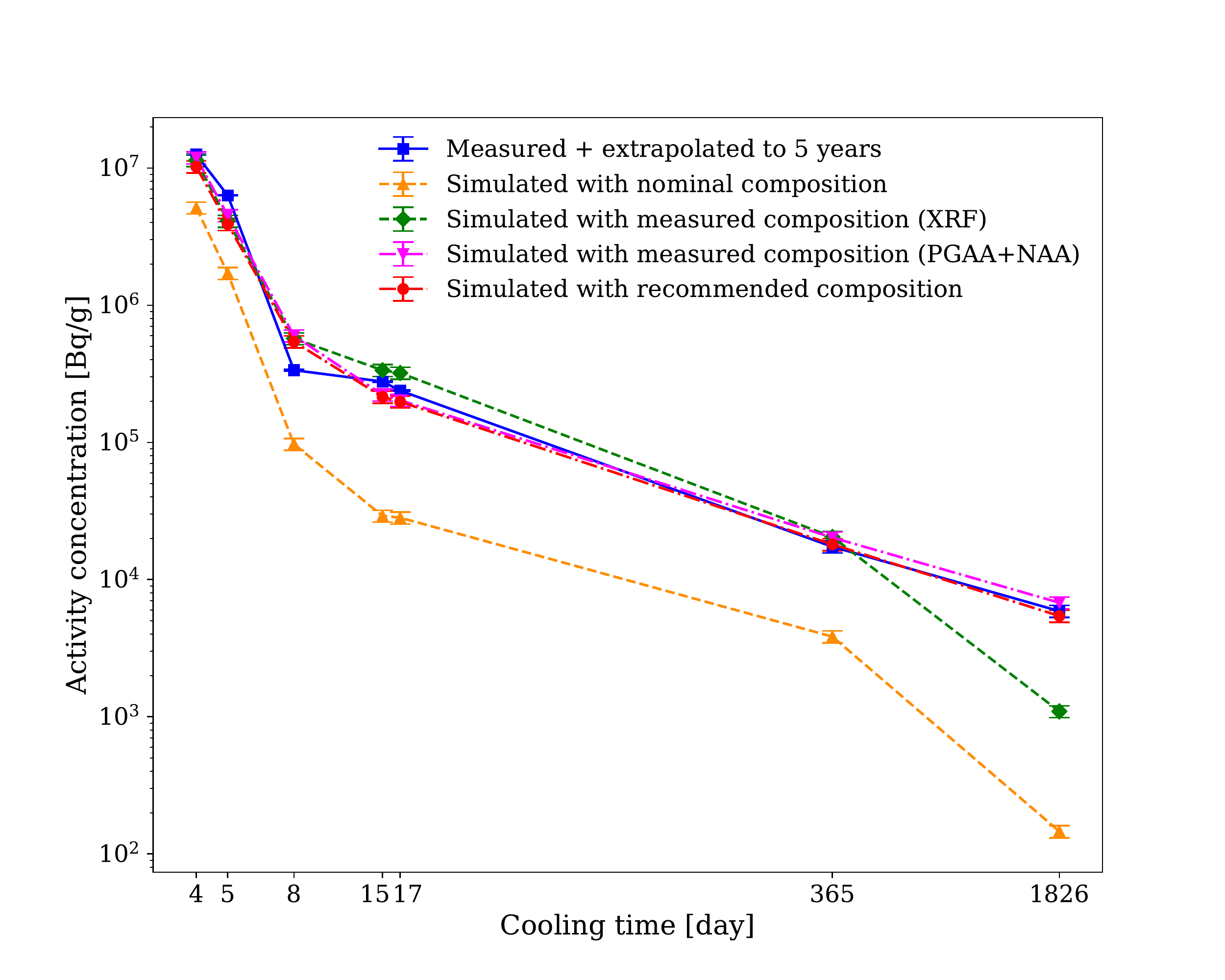}
    \caption{\label{matComp_f_PEB4C}}
  \end{subfigure}
  \caption{Measured decay profiles of PE-B4C-concrete and simulated decay profiles with different initial compositions after irradiation in the Thermalized~(\ref{matComp_th_PEB4C}) and Fast channels~(\ref{matComp_f_PEB4C}). The statistical uncertainties are too small to be discernible. The measured points are connected for better visibility.\label{matComp_PEB4C} }
\end{figure}

\begin{figure}[ht!]
  \centering
   \begin{subfigure}[b]{0.48\textwidth}
    \includegraphics[width=\textwidth]{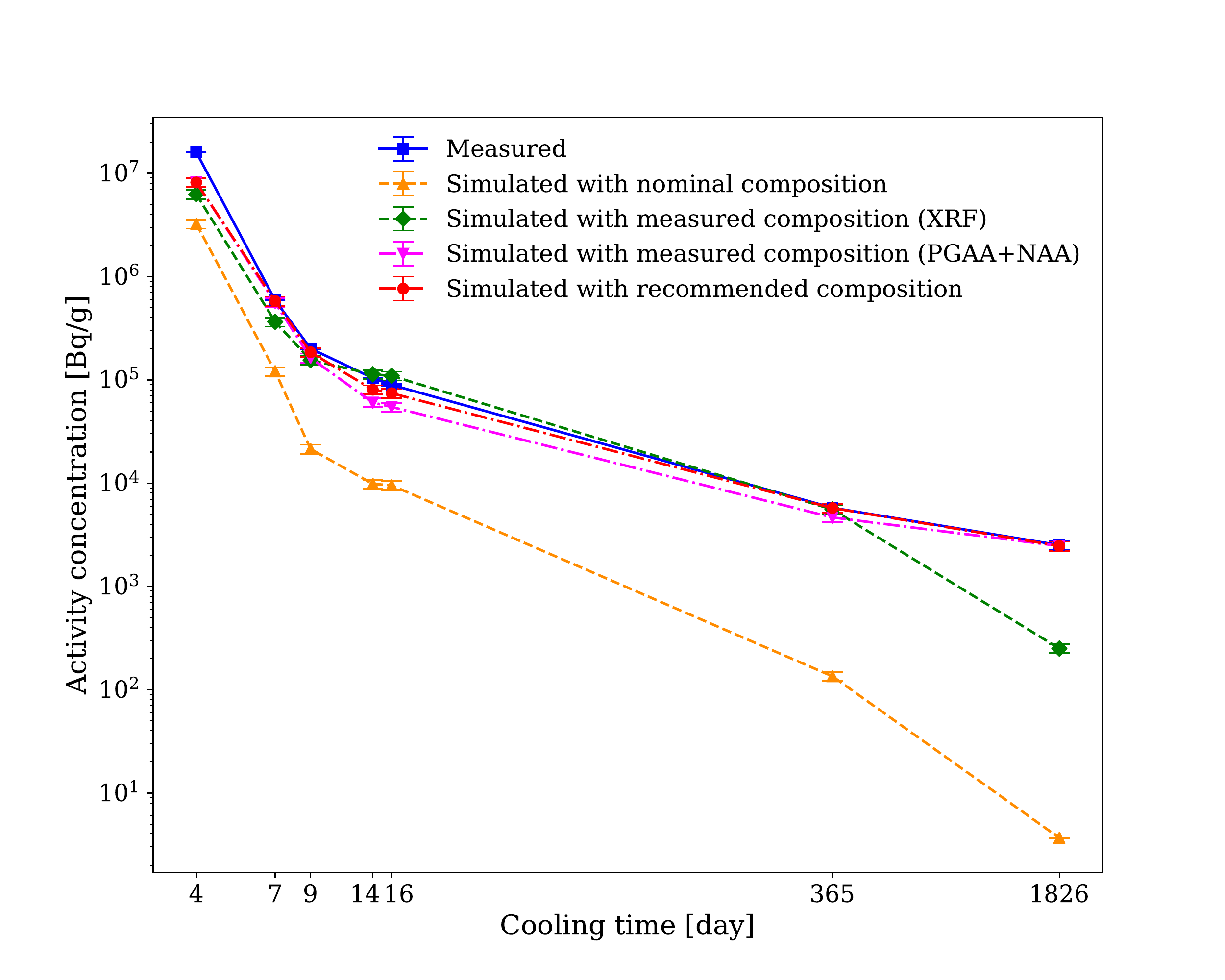}
    \caption{\label{matComp_th_Ref}}
  \end{subfigure}
  ~
  \begin{subfigure}[b]{0.48\textwidth}
    \includegraphics[width=\textwidth]{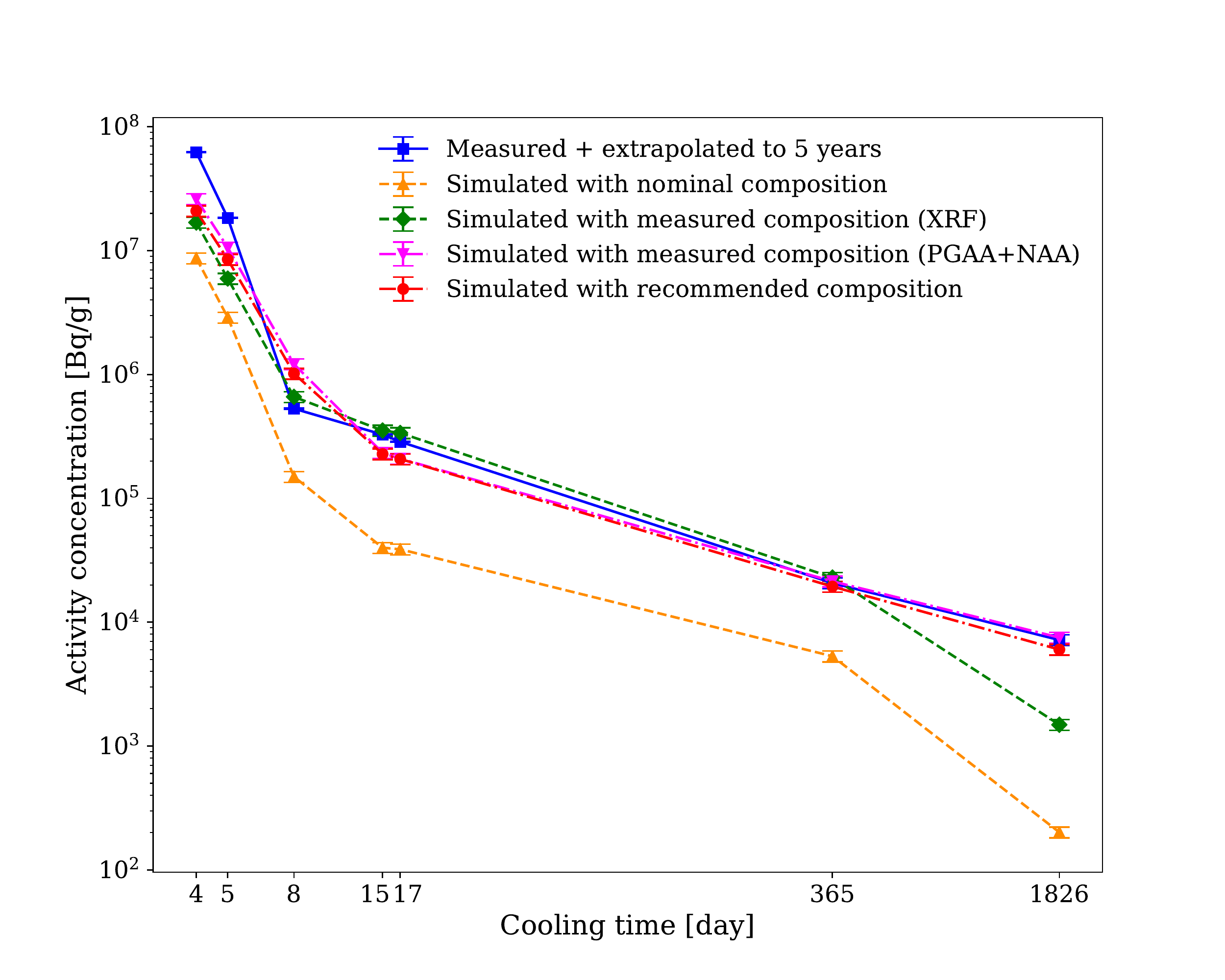}
    \caption{\label{matComp_f_Ref}}
  \end{subfigure}
  \caption{Measured decay profiles of Reference concrete and simulated decay profiles with different initial compositions after irradiation in the Thermalized~(\ref{matComp_th_Ref}) and Fast channels~(\ref{matComp_f_Ref}). The statistical uncertainties are too small to be discernible. The measured points are connected for better visibility.\label{matComp_Ref} }
\end{figure}

\begin{figure}[ht!]
  \centering
   \begin{subfigure}[b]{0.48\textwidth}
    \includegraphics[width=\textwidth]{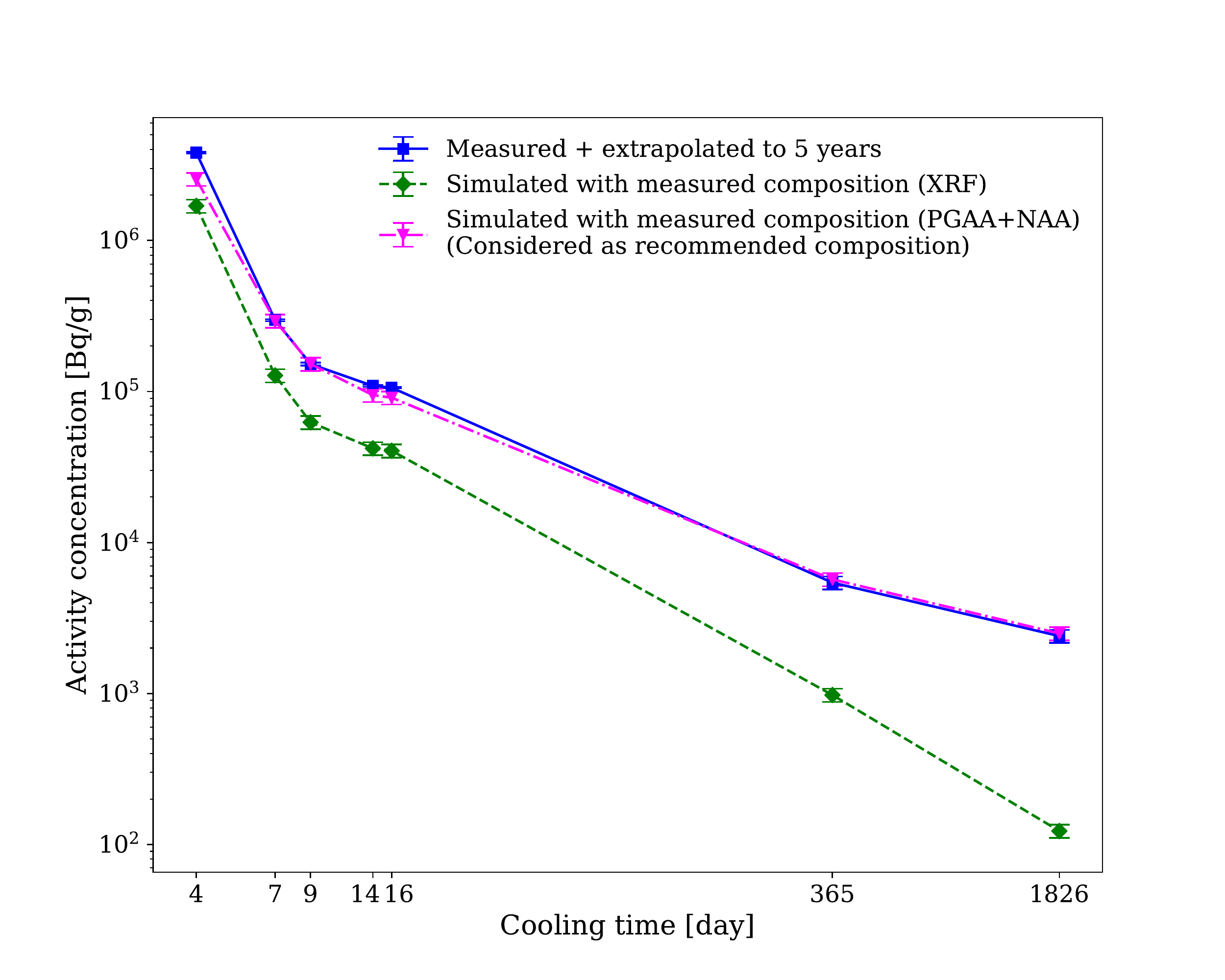}
    \caption{\label{matComp_th_Sk}}
  \end{subfigure}
  ~
  \begin{subfigure}[b]{0.48\textwidth}
    \includegraphics[width=\textwidth]{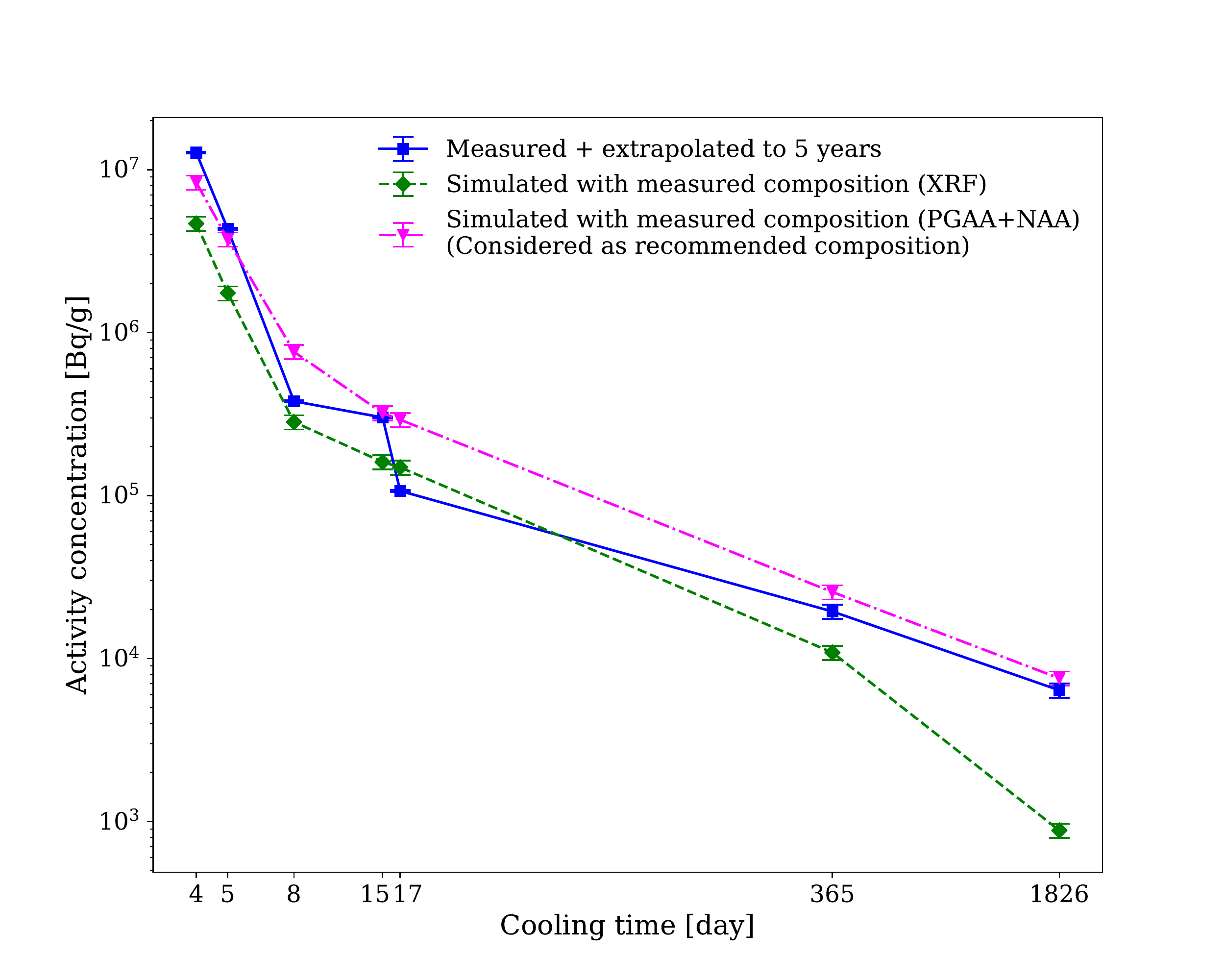}
    \caption{\label{matComp_f_Sk}}
  \end{subfigure}
  \caption{Measured decay profiles of Skanska concrete and simulated decay profiles with different initial compositions after irradiation in the Thermalized~(\ref{matComp_th_Sk}) and Fast channels~(\ref{matComp_f_Sk}). The statistical uncertainties are too small to be discernible. The measured points are connected for better visibility.\label{matComp_Sk} }
\end{figure}

The 2~hours irradiation of the samples in the Thermalized and Fast channels of the BRR were reproduced via MCNPX simulations applied in combination with Cinder1.05 analytical activation code. The measured and simulated activity concentrations (extrapolated to 1~y and 5~y of cooling in the case of the measured data) are presented in~\Cref{matComp_PEB4C,matComp_Ref,matComp_Sk} for all three concretes. It is revealed that the simulations performed only with the nominal concrete composition significantly underestimated the measured activities. The simulations gave only 0.1--28\% and 10--41\% of the measured activity for PE-B4C-concrete, and 0.2--20\% and 12--28\% for the Reference concrete in the Thermalized and Fast channels, respectively. However, it is shown that the discrepancies between the simulated and measured results are remarkably reduced with the application of either of the measurement-based composition. For all the concretes the first measured activity is consequently underestimated with all initial composition, understandably as the contribution of the key isotopes here is the lowest. Except of this data point, for the Thermalized channel (\Cref{matComp_th_PEB4C,matComp_th_Ref,matComp_th_Sk}) the discrepancies are in average within 40\% based on either analytical method for most of the studied time period. In the case of EDXRF measurement based simulations, the activity after 5~y cooling is underestimated with one order of magnitude, as it was shown that the main contributors to the activity at this time are the $^{60}$Co and $^{152}$Eu, and none of their parent elements were measurable by the EDXRF method.

The results of the Fast channel (\Cref{matComp_f_PEB4C,matComp_f_Ref,matComp_f_Sk}) are more varied, most likely due to the presence of threshold reactions and the resonance integral region, making the simulations particularly sensitive to the uncertainties of the irradiating neutron spectra. However, the  activity simulations with the measured compositions are considerably more realistic for this case too, resulting at least 2 times higher activities than with the nominal composition. In essence, the activity increase due to the trace elements is comparable with the typical two-threefold safety reserve typically imposed on Monte Carlo dose simulations.

\bigskip
Revealing the relevance of the initial concrete composition in the neutron activation simulations, simplified but realistic input compositions (material cards) were prepared for the PE-B4C- and Reference concretes from the combination of all the measured initial compositions focusing on the nominal bulk composition and the parent elements of the key isotopes. The derived compositions, which are recommended for activation simulations are presented as ready-to-use MCNP material cards in Table~\ref{tab:recComp}.


\captionof{table}{MCNP style compositions (material cards) for PE-B4C and Reference concretes, recommended for neutron activation simulations.\label{tab:recComp}}
\begin{boxedminipage}[t]{0.48\textwidth}
  \small
  \setstretch{0.8}
  \texttt{
    \newline
    PE-B4C concrete composition
    \newline
    by weight fraction
    \newline
    Density [g/cm3] = 1.97 \\
    \newline
    \centering
    \begin{tabular}{rc}
      1001 & -1.2633E-02 \\
      5000 & -2.0000E-03 \\
      6000 & -5.4583E-02 \\
      8016 & -4.6618E-01 \\
      11000 & -1.2910E-02 \\
      12000 & -9.8257E-04 \\
      13000 & -5.5120E-02 \\
      14000 & -2.7037E-01 \\
      16000 & -2.3669E-03 \\
      17000 & -1.3000E-04 \\
      19000 & -1.9467E-02 \\
      20000 & -8.5427E-02 \\
      21000 & -3.4450E-06 \\
      22000 & -1.5900E-03 \\
      24000 & -9.4300E-05 \\
      26000 & -1.5870E-02 \\
      27000 & -5.7810E-06 \\
      30000 & -1.0006E-04 \\
      37000 & -6.8500E-05 \\
      55000 & -1.9830E-06 \\
      57000 & -1.9965E-05 \\
      62000 & -1.5670E-06 \\
      63000 & -5.8510E-07 \\
      70000 & -1.0260E-06 \\
      74000 & -7.5460E-05 \\
      90000 & -4.3010E-06 \\
  \end{tabular}}
  \label{box:MatCardPEB4C}
\end{boxedminipage}
~
\begin{boxedminipage}[t]{0.48\textwidth}
  \small
  \setstretch{0.8}
  \texttt{
    \newline
    Reference concrete composition
    \newline
    by weight fraction
    \newline
    Density [g/cm3] = 2.33
    \newline
    \begin{tabular}{rc}
      1001 & -6.0087E-03 \\
      5000 & -2.4900E-05 \\
      8016 & -4.8801E-01 \\
      11000 & -1.9676E-02 \\
      12000 & -9.5302E-03 \\
      13000 & -6.6560E-02 \\
      14000 & -3.0098E-01 \\
      16000 & -2.8035E-03 \\
      17000 & -3.0200E-05 \\
      19000 & -2.1899E-02 \\
      20000 & -6.6338E-02 \\
      21000 & -3.9660E-06 \\
      22000 & -1.7600E-03 \\
      24000 & -6.5710E-05 \\
      26000 & -1.5500E-02 \\
      27000 & -5.5230E-06 \\
      30000 & -1.0680E-04 \\
      37000 & -8.6160E-05 \\
      55000 & -2.5860E-06 \\
      57000 & -2.3716E-05 \\
      62000 & -2.3780E-06 \\
      63000 & -6.9960E-07 \\
      74000 & -5.9570E-04 \\
      90000 & -7.6160E-06 
  \end{tabular}}
  \end{boxedminipage}
\label{box:MatCardRef}

\clearpage
Irradiation simulations were performed with these `recommended compositions' of the PE-B4C- and the Reference concretes as well for the whole studied time period. In addition, the isotopic activities were determined for the t$_{0}$ -- end of irradiation time -- and compared to that of extrapolated from measured data for the same time. As an example, results for the PE-B4C-concrete are presented and compared in Table~\ref{tab:MeasSimIsoPE}. It was found that the activities agree well, within 15\% for most isotopes. High, $>$50\% differences found for the $^{140}$La in the Thermalized irradiation channel, and for $^{54}$Mn and $^{157}$Yb in the Fast channel. The case of the La can be attributed to the fact that the La content was found to be very inhomogeneous among the samples, while Yb, Fe and Mn have a significant resonance integral in the high energy end of the applied neutron spectrum, being particularly  sensitive for spectral uncertainties. These differences highlight the importance of extensive sampling for such inhomogeneous materials like concretes.

\begin{table}[htbp]
  \centering
  \caption{Comparison of simulation- and measurement-based extrapolated isotopic activity concentrations at t$_{0}$ in PE-B4C-concrete.}
  \label{tab:MeasSimIsoPE}
  \resizebox *{\textwidth}{!}{
    \begin{tabular}{lcccccc}
      \hline
      & \multicolumn{3}{c}{Thermalized channel} & \multicolumn{3}{c}{Fast channel} \\
      \hline
      \multirow{2}{*}{Isotope} & \multicolumn{2}{c}{Activity concentration [Bq/g]} & \multirow{2}{*}{Difference [\%]} & \multicolumn{2}{c}{Activity concentration [Bq/g]} &  \multirow{2}{*}{Difference [\%]} \\
      & {measurement} & {simulation} & & {measurement} & {simulation} & \\
      \hline
      {Na-24} &
      {2.65E+08} &
      {3.53E+08} &
      {33} &
      {9.15E+08} &
      {7.59E+08} &
      {{}-17}\\
      {W-187} &
      {4.51E+06} &
      {3.95E+06} &
      {{}-12} &
      {1.80E+07} &
      {1.51E+07} &
      {{}-16}\\
      {La-140} &
      {3.90E+05} &
      {5.96E+05} &
      {53} &
      {1.46E+06} &
      {1.32E+06} &
      {{}-9}\\
      {Sm-153} &
      {4.13E+05} &
      {3.06E+05} &
      {{}-26} &
      {1.92E+06} &
      {1.44E+06} &
      {{}-25}\\
      {Yb-175} &
      {3.46E+04} &
      {2.40E+04} &
      {{}-31} &
      {1.08E+05} &
      {5.00E+04} &
      {{}-54}\\
      {Rb-86} &
      {1.28E+04} &
      {1.50E+04} &
      {17} &
      {7.08E+04} &
      {7.02E+04} &
      {{}-1}\\
      {Pa-233} &
      {4.45E+03} &
      {3.88E+03} &
      {{}-13} &
      {2.24E+04} &
      {1.60E+04} &
      {{}-29}\\
      {Cr-51} &
      {3.16E+04} &
      {3.55E+04} &
      {12} &
      {7.60E+04} &
      {7.60E+04} &
      {0}\\
      {Fe-59} &
      {1.75E+04} &
      {1.63E+04} &
      {{}-7} &
      {4.47E+04} &
      {3.77E+04} &
      {{}-16}\\
      {Sc-46} &
      {1.71E+04} &
      {1.94E+04} &
      {13} &
      {4.61E+04} &
      {4.12E+04} &
      {{}-11}\\
      {Zn-65} &
      {1.85E+03} &
      {1.87E+03} &
      {1} &
      {4.83E+03} &
      {4.76E+03} &
      {{}-1}\\
      {Mn-54} &
      {2.69E+02} &
      {2.89E+02} &
      {7} &
      {8.53E+03} &
      {1.27E+04} &
      {49}\\
      {Cs-134} &
      {6.06E+02} &
      {5.97E+02} &
      {{}-1} &
      {2.69E+03} &
      {2.49E+03} &
      {{}-7}\\
      {Co-60} &
      {1.38E+03} &
      {1.41E+03} &
      {3} &
      {3.76E+03} &
      {3.47E+03} &
      {{}-8}\\
      {Eu-152} &
      {1.60E+03} &
      {1.67E+03} &
      {5} &
      {4.35E+03} &
      {3.58E+03} &
      {{}-18}\\
      \hline
  \end{tabular}}
\end{table}

The total activity concentrations -- from the 15 key isotopes -- were also determined with the recommended concrete compositions for both concretes for the studied time interval, and are presented in \Cref{matComp_PEB4C,matComp_Ref}. The simulations with the recommended compositions gave agreement with the measured activities in the Thermalized channels: agreeing within 38\% and 49\% in the whole time period for the PE-B4C- and the Reference concrete, and within 16\% and 23\% without the obviously underestimated 1st data point at day~4. In the Fast channel the recommended composition-based simulated activities are more similar to the purely measurement-based ones, fortifying that high energy simulations may be more sensitive to other parameters, e.g.\ the uncertainties of the irradiating spectrum.

For the Skanska concrete no specific `recommended composition' was developed as it was not considered as shielding material in the region of interest of the current study, although, in Figure~\ref{matComp_Sk} it is shown that the activation analysis based composition 
can be recommended for simulations as well.

\subsection{Decay gamma dose rate simulations in the ESS bunker}

\begin{figure}[ht!]
  \centering
  \begin{subfigure}[t]{0.39\textwidth}
    \includegraphics[width=\textwidth]{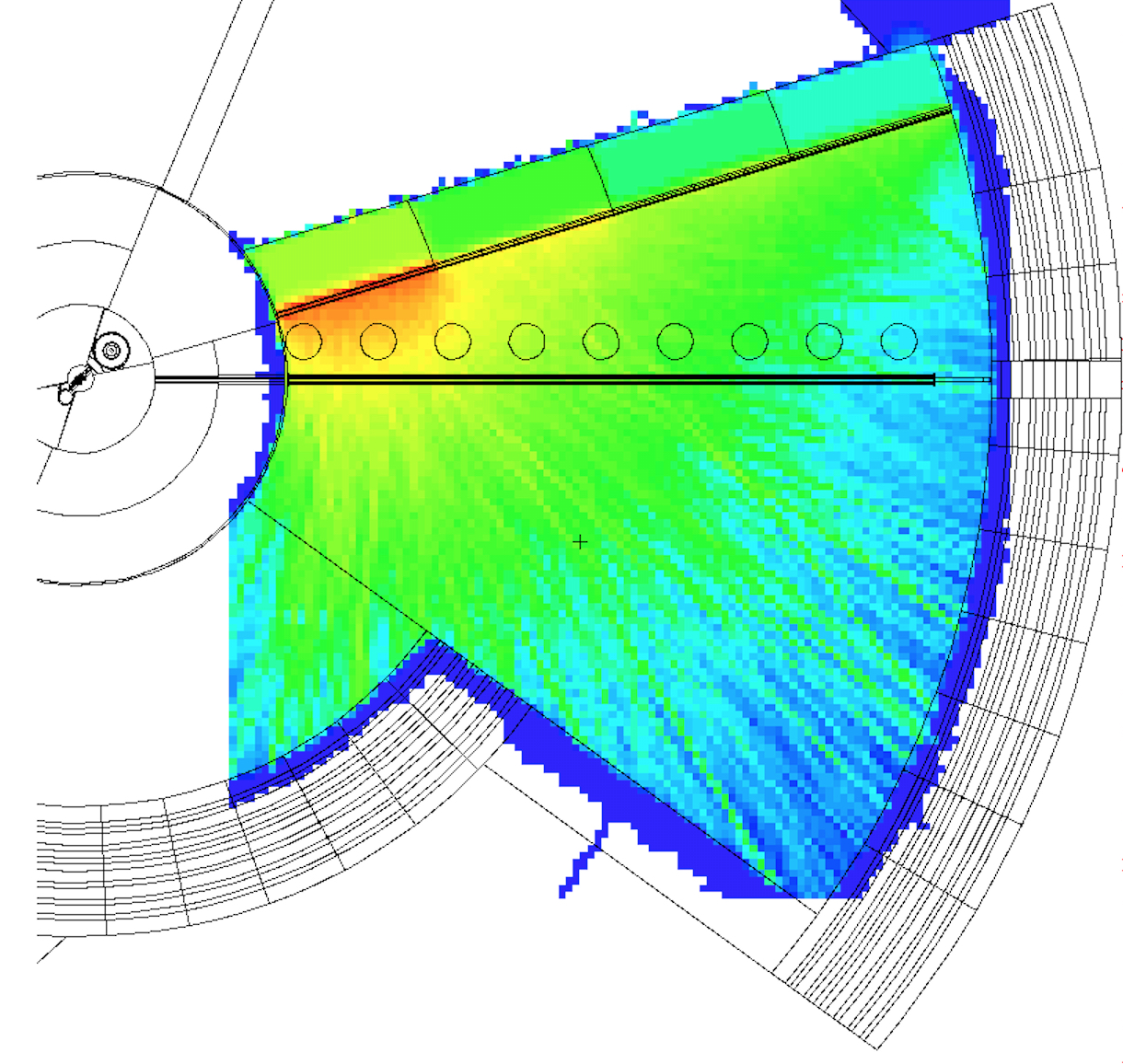}
    \caption{\label{dm_ref}}
  \end{subfigure}
  \begin{subfigure}[t]{0.40\textwidth}
    \includegraphics[width=\textwidth]{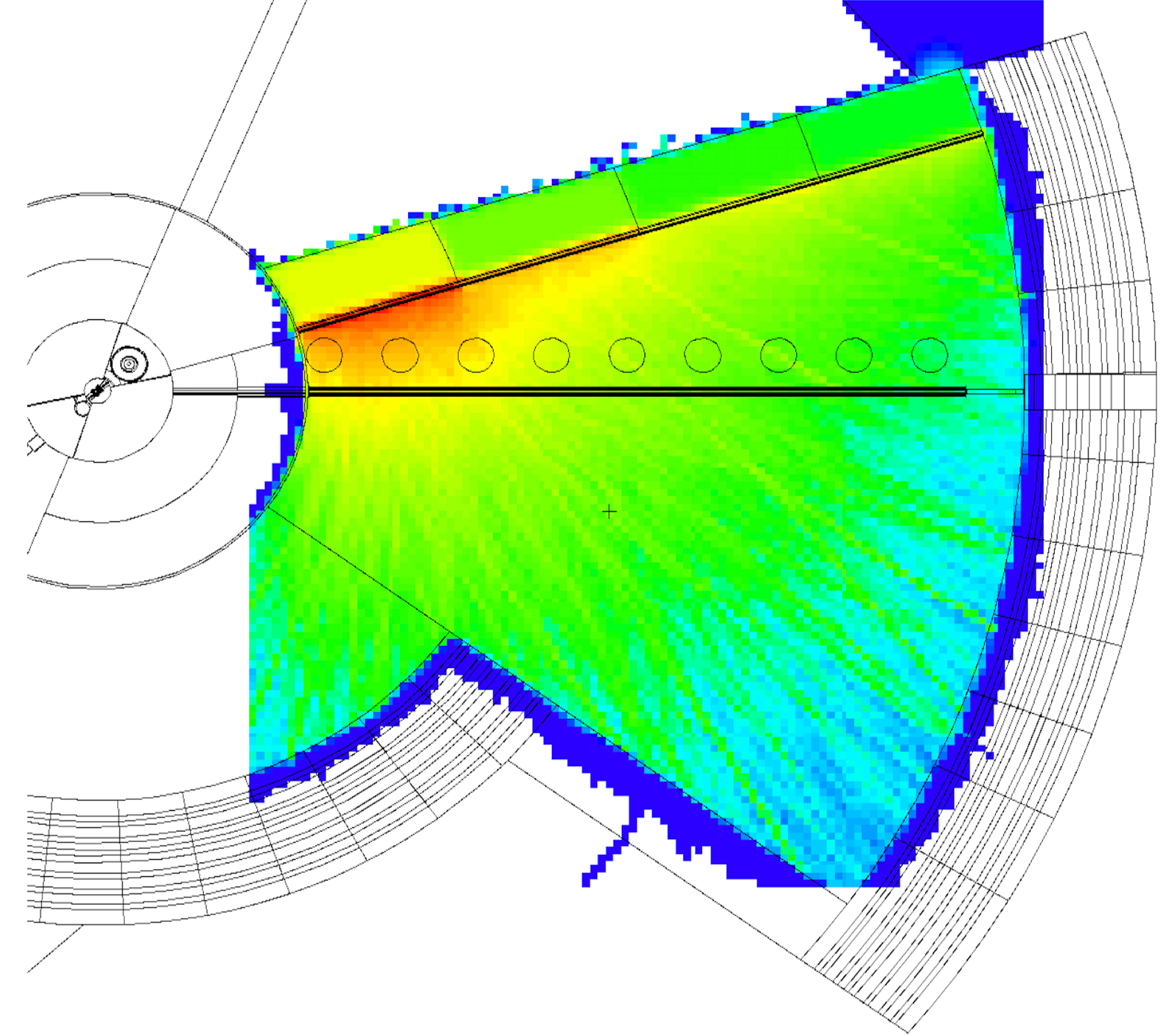}
    \caption{.\label{dm_rec}}
  \end{subfigure}
  \begin{subfigure}[t]{0.15\textwidth}
    \includegraphics[width=\textwidth]{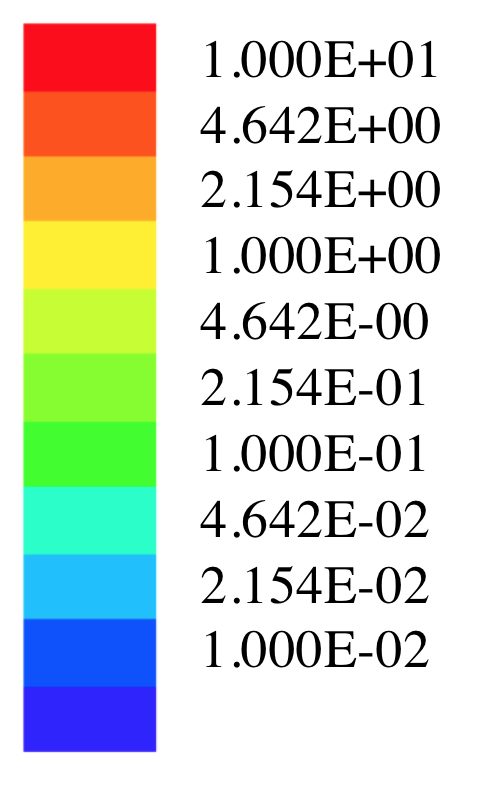}
  \end{subfigure}

  \caption{Simulated dose rate maps [$\rm \frac{\mu Sv}{h} $] from a the segmented axial wall made of Reference concrete. Simulations were performed with nominal~(\ref{dm_ref}) and recommended~(\ref{dm_rec}) initial concrete compositions. \label{DoseMap} }
\end{figure}

The power of the developed realistic `recommended compositions' were studied not only for activation, but in terms of dosimetric simulations as well, for a single radial wall in the West Hall of the ESS bunker. Decay gamma dose rates were determined in the Hall, generated by the activated isotopes of the wall in a typical maintenance scenario with 3 days of cooling after the beam shut down. The simulated dose maps for both compositions are shown in Figure~\ref{DoseMap}, and the dose rates are numerically compared in nine locations along the BIFROST instrument, as presented in Table~\ref{tab:DosePoints}.

In Figure~\ref{DoseMap} it is demonstrated that the dose rates significantly increase in the close proximity of the first quarter of the axial wall, and the area with $>$1~$\rm \frac{\mu Sv}{h}$ gamma dose rate (yellow) is also increased, due to the trace elements in one single wall of the bunker.  

As the gamma dose rate is quite inhomogeneous in the bunker, the increase of dose rate in the selected areas also varies between 29--72\%, exceeding 50\% in five of nine positions sampled for numerical comparison, as described in Figure~\ref{bunker_geom}. It was revealed that the presence or lack of trace elements in these simulations by itself can multiply the  simulated activities and doses, depleting the typical conservative allowance of Monte Carlo safety simulations. Herewith the need for more realistic material compositions for activation and safety simulations is confirmed.

\begin{table}[htbp]
  \centering
  \caption{Simulated dose rates [$\rm \frac{\mu Sv}{h} $]  in the West Hall of the ESS bunker. Locations are indicated in Fig.~\ref{bunker_geom}. }
  \label{tab:DosePoints}
  \resizebox{0.75\textwidth}{!}{
    \begin{tabular}{lccccc}
      \hline
      & \multicolumn{2}{c}{Nominal composition} &  \multicolumn{2}{c}{Recommended composition} &  \\
      \hline
      & Dose rate & Unc. [\%] & Dose rate & Unc. [\%] & Increase [\%] \\
      \hline
      1 & 1.89E+00 &  7 & 2.73E+00 &  6 & 45 \\
      2 & 1.79E+00 &  8 & 2.37E+00 &  6 & 33 \\
      3 & 9.37E-01 &  9 & 1.32E+00 &  7 & 41 \\
      4 & 4.99E-01 & 11 & 6.45E-01 &  7 & 29 \\
      5 & 2.52E-01 &  9 & 4.02E-01 &  7 & 59 \\
      6 & 1.87E-01 &  9 & 3.01E-01 &  7 & 61 \\
      7 & 1.24E-01 & 10 & 2.14E-01 &  8 & 72 \\
      8 & 7.75E-02 & 12 & 1.33E-01 &  8 & 72 \\
      9 & 6.07E-02 & 27 & 1.04E-01 & 17 & 71 \\
      \hline
  \end{tabular}}
\end{table}

\section{Conclusions}

A comprehensive study was carried out on the activation properties of neutron shielding concretes. Neutron activation of three different concretes that were considered 
to be used at ESS: PE-B4C-concrete, its Reference concrete and a Skanska concrete were measured in the Thermalized and Fast channels of the BRR. Following the decrease of activity of the irradiated concrete samples in a 2~weeks period, it was found that the overall activity concentration of the PE-B4C-concrete is significantly lower than that of the Reference concrete: 54\% and 66\% lower during the first week of cooling after being irradiated in the Thermalized and Fast channels, respectively. Thus it is proven that the 
PE-B4C-concrete is more advantageous than the Reference concrete in terms of activation, especially during the first week of cooling, i.e.\ in a potential maintenance period.

In order to take into account the activation products more realistically in Monte Carlo simulations, the composition of all three concretes were determined with three analytical methods (NAA, PGAA, XRF). On the basis of the performed neutron activation measurements, the 15 most important gamma-emitting radioisotopes -- giving together a minimum of 83\% of the total measured activity for the whole studied time range -- were identified, allowing a simple comparison of the measured and simulated activities. This simplification can be used in activation related safety considerations to evaluate both short- and long-term effects, not only in ESS, but also in other facilities associated with neutron irradiation (e.g.\ in nuclear power plants). 

Activation simulations of the three concretes were performed with MCNP and Cinder1.05 codes for all the measured and nominal compositions (similar to currently available general material cards). It was shown that the simulations with the nominal composition underestimate the measured activities with one order of magnitude, highlighting the importance of trace elements in terms of neutron activation. On the basis of the measured compositions, MCNP material cards were developed and validated for the PE-B4C- and Reference concretes, that are recommended for activation calculations, reproducing the measured data mostly 
within 23\% difference in the Thermalized channel.


The importance of accurate material cards was also demonstrated in terms of radiation safety: decay gamma dose rates were determined for a single concrete wall of the ESS West Hall with realistic irradiation scenario for the Reference concrete. It was found that after 3 days of cooling i.e.\ in a maintenance period, the decay gamma dose rates simulated with the recommended, measurement-based material card resulted 29-72\% higher dose rates than with the nominal composition which does not contain trace elements.

To sum up, a detailed, comprehensive study was performed on the neutron activation properties of neutron shielding concretes and the measurement-based methodology was presented for the compilation of detailed material cards for more realistic activation simulations.

\section*{Acknowledgments}

This project has received funding from the European Union’s Horizon 2020 research and innovation program under grant agreement No 654000. Computing resources were provided by DMSC Computing Centre (https://europeanspallationsource.se/data-management-software/computing-centre).


\section*{Appendix}
~
\setcounter{table}{0}
\renewcommand{\thetable}{A\arabic{table}}
~
\begin{table}[htbp]
  \centering
  \caption{Measured elemental composition of PE-B4C-concrete with PGAA, XRF (most of them with EDXRF, marked (*) ones with WDXRF) and NAA analytical methods. }
  \label{tab:compPE}
  \resizebox *{!}{\textheight}{
\begin{tabular}{cccccccc}
      \hline
      Element  &  Nominal  &  PGAA  &  Unc. [\%]  &  XRF  &  Unc. [\%]  &  NAA  &  Unc. [\%]  \\
      \hline
        &  \multicolumn{7}{c}{ [w\%]}  \\
      \hline 
H &
{\selectlanguage{english} 2.31} &
{\selectlanguage{english} 1.26} &
{\selectlanguage{english} 3.2} &
{\selectlanguage{english} {}-} &
{\selectlanguage{english}\sffamily \foreignlanguage{english}{\textrm{{}-~}}} &
{\selectlanguage{english} {}-} &
{\selectlanguage{english}\sffamily \foreignlanguage{english}{\textrm{{}-~}}}\\
{\selectlanguage{english} C} &
{\selectlanguage{english} 8.99} &
{\selectlanguage{english} 5.46} &
{\selectlanguage{english} 11.9} &
{\selectlanguage{english} {}-} &
{\selectlanguage{english}\sffamily \foreignlanguage{english}{\textrm{{}-~}}} &
{\selectlanguage{english} {}-} &
{\selectlanguage{english}\sffamily \foreignlanguage{english}{\textrm{{}-~}}}\\
{\selectlanguage{english} Na} &
{\selectlanguage{english} 0.62} &
{\selectlanguage{english} 1.11} &
{\selectlanguage{english} 4.2} &
{\selectlanguage{english} 1.29} &
{\selectlanguage{english} 15.2} &
{\selectlanguage{english} 1.21} &
{\selectlanguage{english} 3.7}\\
{\selectlanguage{english} Al} &
{\selectlanguage{english} 2.35} &
{\selectlanguage{english} 3.44} &
{\selectlanguage{english} 3.7} &
{\selectlanguage{english} 5.51} &
{\selectlanguage{english} 0.6} &
{\selectlanguage{english} {}-} &
{\selectlanguage{english}\sffamily \foreignlanguage{english}{\textrm{{}-~}}}\\
{\selectlanguage{english} Si} &
{\selectlanguage{english} 28.6} &
{\selectlanguage{english} 21.5} &
{\selectlanguage{english} 3.3} &
{\selectlanguage{english} 27.0} &
{\selectlanguage{english}\sffamily \foreignlanguage{english}{\textrm{0.1}}} &
{\selectlanguage{english} {}-} &
{\selectlanguage{english}\sffamily \foreignlanguage{english}{\textrm{{}-~}}}\\
{\selectlanguage{english} S} &
{\selectlanguage{english} 0.28} &
{\selectlanguage{english} 0.23} &
{\selectlanguage{english} 5.5} &
{\selectlanguage{english} 0.24} &
{\selectlanguage{english} 6.5} &
{\selectlanguage{english} {}-} &
{\selectlanguage{english}\sffamily \foreignlanguage{english}{\textrm{{}-~}}}\\
{\selectlanguage{english} K} &
{\selectlanguage{english} 1.26} &
{\selectlanguage{english} 1.39} &
{\selectlanguage{english} 3.7} &
{\selectlanguage{english} 1.95} &
{\selectlanguage{english}\sffamily \foreignlanguage{english}{\textrm{0.1}}} &
{\selectlanguage{english} {}-} &
{\selectlanguage{english}\sffamily \foreignlanguage{english}{\textrm{{}-~}}}\\
{\selectlanguage{english} Ca} &
{\selectlanguage{english} 8.10} &
{\selectlanguage{english}\sffamily \foreignlanguage{english}{\textrm{6.50}}} &
{\selectlanguage{english} 3.9} &
{\selectlanguage{english} 8.54} &
{\selectlanguage{english}\sffamily \foreignlanguage{english}{\textrm{0.1}}} &
{\selectlanguage{english} {}-} &
{\selectlanguage{english}\sffamily \foreignlanguage{english}{\textrm{{}-~}}}\\
{\selectlanguage{english} Fe} &
{\selectlanguage{english} 0.84} &
{\selectlanguage{english} 1.12} &
{\selectlanguage{english} 1.6} &
{\selectlanguage{english} 1.47} &
{\selectlanguage{english}\sffamily \foreignlanguage{english}{\textrm{0.1}}} &
{\selectlanguage{english} 1.59} &
{\selectlanguage{english}\sffamily \foreignlanguage{english}{\textrm{4.6}}}\\
{\selectlanguage{english} B} &
{\selectlanguage{english} 0.6} &
{\selectlanguage{english} 0.2} &
{\selectlanguage{english} 20} &
{\selectlanguage{english} {}-} &
{\selectlanguage{english}\sffamily \foreignlanguage{english}{\textrm{{}-~}}} &
{\selectlanguage{english} {}-} &
{\selectlanguage{english}\sffamily \foreignlanguage{english}{\textrm{~-}}}\\
{\selectlanguage{english} O} &
{\selectlanguage{english} 45.8} &
{\selectlanguage{english}\sffamily \foreignlanguage{english}{\textrm{56.4}}} &
{\selectlanguage{english} 4} &
{\selectlanguage{english}\sffamily \foreignlanguage{english}{\textrm{~-}}} &
{\selectlanguage{english}\sffamily \foreignlanguage{english}{\textrm{{}-~}}} &
{\selectlanguage{english}\sffamily \foreignlanguage{english}{\textrm{~-}}} &
{\selectlanguage{english}\sffamily \foreignlanguage{english}{\textrm{~-}}}\\
      \hline
       &  \multicolumn{7}{c}{  [mg/kg] }  \\  
      \hline 
{\selectlanguage{english} Mg} &
{\selectlanguage{english} 1960} &
{\selectlanguage{english} {}-} &
{\selectlanguage{english}\sffamily \foreignlanguage{english}{\textrm{{}-~}}} &
{\selectlanguage{english}\sffamily \foreignlanguage{english}{\textrm{983}}} &
{\selectlanguage{english} 30} &
{\selectlanguage{english} {}-} &
{\selectlanguage{english}\sffamily \foreignlanguage{english}{\textrm{~-}}}\\
{\selectlanguage{english} Ti} &
{\selectlanguage{english} 521} &
{\selectlanguage{english} 1030} &
{\selectlanguage{english} 1.7} &
{\selectlanguage{english} 1590} &
{\selectlanguage{english} 0.6} &
{\selectlanguage{english} {}-} &
{\selectlanguage{english}\sffamily \foreignlanguage{english}{\textrm{~-}}}\\
{\selectlanguage{english} P} &
{\selectlanguage{english} 260} &
{\selectlanguage{english} {}-} &
{\selectlanguage{english}\sffamily \foreignlanguage{english}{\textrm{{}-~}}} &
{\selectlanguage{english} {}-} &
{\selectlanguage{english}\sffamily \foreignlanguage{english}{\textrm{{}-~}}} &
{\selectlanguage{english} {}-} &
{\selectlanguage{english}\sffamily \foreignlanguage{english}{\textrm{~-}}}\\
{\selectlanguage{english} Cl} &
{\selectlanguage{english} 36} &
{\selectlanguage{english} 156} &
{\selectlanguage{english} 2.7} &
{\selectlanguage{english}\sffamily \foreignlanguage{english}{\textrm{130}}} &
{\selectlanguage{english} 7.7} &
{\selectlanguage{english} {}-} &
{\selectlanguage{english}\sffamily \foreignlanguage{english}{\textrm{~-}}}\\
{\selectlanguage{english} V} &
{\selectlanguage{english}\sffamily \foreignlanguage{english}{\textrm{{}-~}}} &
{\selectlanguage{english} 77} &
{\selectlanguage{english} 6} &
{\selectlanguage{english} 57.6} &
{\selectlanguage{english} 12.3} &
{\selectlanguage{english} {}-} &
{\selectlanguage{english}\sffamily \foreignlanguage{english}{\textrm{~-}}}\\
{\selectlanguage{english} Sm} &
{\selectlanguage{english}\sffamily \foreignlanguage{english}{\textrm{{}-~}}} &
{\selectlanguage{english} 1.5} &
{\selectlanguage{english} 6} &
{\selectlanguage{english} {}-} &
{\selectlanguage{english}\sffamily \foreignlanguage{english}{\textrm{{}-~}}} &
{\selectlanguage{english} 1.57} &
{\selectlanguage{english} 4.1}\\
{\selectlanguage{english} Gd} &
{\selectlanguage{english}\sffamily \foreignlanguage{english}{\textrm{{}-~}}} &
{\selectlanguage{english} 1.6} &
{\selectlanguage{english} 10} &
{\selectlanguage{english} {}-} &
{\selectlanguage{english}\sffamily \foreignlanguage{english}{\textrm{{}-~}}} &
{\selectlanguage{english} {}-} &
{\selectlanguage{english}\sffamily \foreignlanguage{english}{\textrm{~-}}}\\
{\selectlanguage{english} Mn} &
{\selectlanguage{english}\sffamily \foreignlanguage{english}{\textrm{{}-~}}} &
{\selectlanguage{english} {}-} &
{\selectlanguage{english}\sffamily \foreignlanguage{english}{\textrm{{}-~}}} &
{\selectlanguage{english} 230} &
{\selectlanguage{english} 4.3} &
{\selectlanguage{english} {}-} &
{\selectlanguage{english}\sffamily \foreignlanguage{english}{\textrm{~-}}}\\
{\selectlanguage{english} Sc} &
{\selectlanguage{english}\sffamily \foreignlanguage{english}{\textrm{{}-~}}} &
{\selectlanguage{english} {}-} &
{\selectlanguage{english} {}-} &
{\selectlanguage{english}\sffamily \foreignlanguage{english}{\textrm{9.3*}}} &
{\selectlanguage{english} 29*} &
{\selectlanguage{english} 3.45} &
{\selectlanguage{english} 3.7}\\
{\selectlanguage{english} Cr} &
{\selectlanguage{english}\sffamily \foreignlanguage{english}{\textrm{{}-~}}} &
{\selectlanguage{english} {}-} &
{\selectlanguage{english} {}-} &
{\selectlanguage{english} 82.1} &
{\selectlanguage{english} 1.6} &
{\selectlanguage{english} 94.3} &
{\selectlanguage{english} 4.0}\\
{\selectlanguage{english} Co} &
{\selectlanguage{english}\sffamily \foreignlanguage{english}{\textrm{{}-~}}} &
{\selectlanguage{english} {}-} &
{\selectlanguage{english} {}-} &
{\selectlanguage{english} {}7.6*} &
{\selectlanguage{english}\sffamily \foreignlanguage{english}{\textrm{40*}}} &
{\selectlanguage{english} 5.78} &
{\selectlanguage{english} 4.3}\\
{\selectlanguage{english} Ni} &
{\selectlanguage{english}\sffamily \foreignlanguage{english}{\textrm{{}-~}}} &
{\selectlanguage{english} {}-} &
{\selectlanguage{english} {}-} &
{\selectlanguage{english}\sffamily \foreignlanguage{english}{\textrm{11.7}}} &
{\selectlanguage{english} 6.7} &
{\selectlanguage{english} {}-} &
{\selectlanguage{english}\sffamily \foreignlanguage{english}{\textrm{~-}}}\\
{\selectlanguage{english} Cu} &
{\selectlanguage{english}\sffamily \foreignlanguage{english}{\textrm{{}-~}}} &
{\selectlanguage{english} {}-} &
{\selectlanguage{english} {}-} &
{\selectlanguage{english} 43.5} &
{\selectlanguage{english} 1.5} &
{\selectlanguage{english} {}-} &
{\selectlanguage{english}\sffamily \foreignlanguage{english}{\textrm{~-}}}\\
{\selectlanguage{english} Zn} &
{\selectlanguage{english}\sffamily \foreignlanguage{english}{\textrm{{}-~}}} &
{\selectlanguage{english} {}-} &
{\selectlanguage{english} {}-} &
{\selectlanguage{english}\sffamily \foreignlanguage{english}{\textrm{100}}} &
{\selectlanguage{english} 0.9} &
{\selectlanguage{english} 107} &
{\selectlanguage{english} 5.8}\\
{\selectlanguage{english} Ga} &
{\selectlanguage{english}\sffamily \foreignlanguage{english}{\textrm{{}-~}}} &
{\selectlanguage{english} {}-} &
{\selectlanguage{english} {}-} &
{\selectlanguage{english}\sffamily \foreignlanguage{english}{\textrm{7.84}}} &
{\selectlanguage{english} 5.5} &
{\selectlanguage{english} {}-} &
{\selectlanguage{english}\sffamily \foreignlanguage{english}{\textrm{~-}}}\\
{\selectlanguage{english} Ge} &
{\selectlanguage{english}\sffamily \foreignlanguage{english}{\textrm{{}-~}}} &
{\selectlanguage{english} {}-} &
{\selectlanguage{english} {}-} &
{\selectlanguage{english} 4.87} &
{\selectlanguage{english} 5.8} &
{\selectlanguage{english} {}-} &
{\selectlanguage{english}\sffamily \foreignlanguage{english}{\textrm{~-}}}\\
{\selectlanguage{english} As} &
{\selectlanguage{english}\sffamily \foreignlanguage{english}{\textrm{{}-~}}} &
{\selectlanguage{english} {}-} &
{\selectlanguage{english} {}-} &
{\selectlanguage{english} 2.92} &
{\selectlanguage{english} 4.9} &
{\selectlanguage{english} {}-} &
{\selectlanguage{english}\sffamily \foreignlanguage{english}{\textrm{~-}}}\\
{\selectlanguage{english} Rb} &
{\selectlanguage{english}\sffamily \foreignlanguage{english}{\textrm{{}-~}}} &
{\selectlanguage{english} {}-} &
{\selectlanguage{english} {}-} &
{\selectlanguage{english}\sffamily \foreignlanguage{english}{\textrm{59.4}}} &
{\selectlanguage{english} 0.7} &
{\selectlanguage{english} 68.5} &
{\selectlanguage{english} 7.1}\\
{\selectlanguage{english} Sr} &
{\selectlanguage{english}\sffamily \foreignlanguage{english}{\textrm{{}-~}}} &
{\selectlanguage{english} {}-} &
{\selectlanguage{english} {}-} &
{\selectlanguage{english} 311} &
{\selectlanguage{english} 0.4} &
{\selectlanguage{english} {}-} &
{\selectlanguage{english}\sffamily \foreignlanguage{english}{\textrm{~-}}}\\
{\selectlanguage{english} Y} &
{\selectlanguage{english}\sffamily \foreignlanguage{english}{\textrm{{}-~}}} &
{\selectlanguage{english} {}-} &
{\selectlanguage{english} {}-} &
{\selectlanguage{english}\sffamily \foreignlanguage{english}{\textrm{11.6}}} &
{\selectlanguage{english} 1.7} &
{\selectlanguage{english} {}-} &
{\selectlanguage{english}\sffamily \foreignlanguage{english}{\textrm{~-}}}\\
{\selectlanguage{english} Zr} &
{\selectlanguage{english}\sffamily \foreignlanguage{english}{\textrm{{}-~}}} &
{\selectlanguage{english} {}-} &
{\selectlanguage{english} {}-} &
{\selectlanguage{english} 91.4} &
{\selectlanguage{english} 0.4} &
{\selectlanguage{english} {}-} &
{\selectlanguage{english}\sffamily \foreignlanguage{english}{\textrm{~-}}}\\
{\selectlanguage{english}\sffamily \foreignlanguage{english}{\textrm{Nb}}} &
{\selectlanguage{english}\sffamily \foreignlanguage{english}{\textrm{{}-~}}} &
{\selectlanguage{english} {}-} &
{\selectlanguage{english} {}-} &
{\selectlanguage{english} 5.36} &
{\selectlanguage{english} 2.5} &
{\selectlanguage{english} {}-} &
{\selectlanguage{english}\sffamily \foreignlanguage{english}{\textrm{~-}}}\\
{\selectlanguage{english} Mo} &
{\selectlanguage{english}\sffamily \foreignlanguage{english}{\textrm{{}-~}}} &
{\selectlanguage{english} {}-} &
{\selectlanguage{english} {}-} &
{\selectlanguage{english} 3.03} &
{\selectlanguage{english} 4.3} &
{\selectlanguage{english} {}-} &
{\selectlanguage{english}\sffamily \foreignlanguage{english}{\textrm{~-}}}\\
{\selectlanguage{english} In} &
{\selectlanguage{english}\sffamily \foreignlanguage{english}{\textrm{{}-~}}} &
{\selectlanguage{english} {}-} &
{\selectlanguage{english} {}-} &
{\selectlanguage{english}\sffamily \foreignlanguage{english}{\textrm{1.70}}} &
{\selectlanguage{english} 11.4} &
{\selectlanguage{english} {}-} &
{\selectlanguage{english}\sffamily \foreignlanguage{english}{\textrm{~-}}}\\
{\selectlanguage{english} Sn} &
{\selectlanguage{english}\sffamily \foreignlanguage{english}{\textrm{{}-~}}} &
{\selectlanguage{english} {}-} &
{\selectlanguage{english} {}-} &
{\selectlanguage{english} 2.81} &
{\selectlanguage{english} 1.7} &
{\selectlanguage{english} {}-} &
{\selectlanguage{english}\sffamily \foreignlanguage{english}{\textrm{~-}}}\\
{\selectlanguage{english} Sb} &
{\selectlanguage{english}\sffamily \foreignlanguage{english}{\textrm{{}-~}}} &
{\selectlanguage{english} {}-} &
{\selectlanguage{english} {}-} &
{\selectlanguage{english} 1.59} &
{\selectlanguage{english} 14.4} &
{\selectlanguage{english} 1.43} &
{\selectlanguage{english} 8.9}\\
{\selectlanguage{english} Cs} &
{\selectlanguage{english}\sffamily \foreignlanguage{english}{\textrm{{}-~}}} &
{\selectlanguage{english} {}-} &
{\selectlanguage{english} {}-} &
{\selectlanguage{english} 2.58} &
{\selectlanguage{english} 11.7} &
{\selectlanguage{english} 1.98} &
{\selectlanguage{english} 6.5}\\
{\selectlanguage{english} Ba} &
{\selectlanguage{english}\sffamily \foreignlanguage{english}{\textrm{{}-~}}} &
{\selectlanguage{english} {}-} &
{\selectlanguage{english} {}-} &
{\selectlanguage{english}\sffamily \foreignlanguage{english}{\textrm{509}}} &
{\selectlanguage{english} 0.2} &
{\selectlanguage{english} 514} &
{\selectlanguage{english} 4.9}\\
{\selectlanguage{english} La} &
{\selectlanguage{english}\sffamily \foreignlanguage{english}{\textrm{{}-~}}} &
{\selectlanguage{english} {}-} &
{\selectlanguage{english} {}-} &
{\selectlanguage{english} 20.0} &
{\selectlanguage{english} 1.4} &
{\selectlanguage{english}\sffamily \foreignlanguage{english}{\textrm{13.0}}} &
{\selectlanguage{english} 4.0}\\
{\selectlanguage{english} Ce} &
{\selectlanguage{english}\sffamily \foreignlanguage{english}{\textrm{{}-~}}} &
{\selectlanguage{english} {}-} &
{\selectlanguage{english} {}-} &
{\selectlanguage{english} 41.16} &
{\selectlanguage{english} 1.3} &
{\selectlanguage{english}\sffamily \foreignlanguage{english}{\textrm{31.4}}} &
{\selectlanguage{english} 4.2}\\
{\selectlanguage{english}\sffamily \foreignlanguage{english}{\textrm{Pr}}} &
{\selectlanguage{english}\sffamily \foreignlanguage{english}{\textrm{{}-~}}} &
{\selectlanguage{english} {}-} &
{\selectlanguage{english} {}-} &
{\selectlanguage{english} 4.84} &
{\selectlanguage{english} 9.5} &
{\selectlanguage{english} {}-} &
{\selectlanguage{english}\sffamily \foreignlanguage{english}{\textrm{~-}}}\\
{\selectlanguage{english} Nd} &
{\selectlanguage{english}\sffamily \foreignlanguage{english}{\textrm{{}-~}}} &
{\selectlanguage{english} {}-} &
{\selectlanguage{english} {}-} &
{\selectlanguage{english}\sffamily \foreignlanguage{english}{\textrm{19.6}}} &
{\selectlanguage{english} 1.1} &
{\selectlanguage{english}\sffamily \foreignlanguage{english}{\textrm{14.2}}} &
{\selectlanguage{english} 10}\\
{\selectlanguage{english} W} &
{\selectlanguage{english}\sffamily \foreignlanguage{english}{\textrm{{}-~}}} &
{\selectlanguage{english} {}-} &
{\selectlanguage{english} {}-} &
{\selectlanguage{english} {}-} &
{\selectlanguage{english}\sffamily \foreignlanguage{english}{\textrm{{}-~}}} &
{\selectlanguage{english}\sffamily \foreignlanguage{english}{\textrm{75.5}}} &
{\selectlanguage{english} 4.1}\\
{\selectlanguage{english} Pb} &
{\selectlanguage{english}\sffamily \foreignlanguage{english}{\textrm{{}-~}}} &
{\selectlanguage{english} {}-} &
{\selectlanguage{english} {}-} &
{\selectlanguage{english} 18.3} &
{\selectlanguage{english} 4.7} &
{\selectlanguage{english} {}-} &
{\selectlanguage{english}\sffamily \foreignlanguage{english}{\textrm{~-}}}\\
{\selectlanguage{english} Th} &
{\selectlanguage{english}\sffamily \foreignlanguage{english}{\textrm{{}-~}}} &
{\selectlanguage{english} {}-} &
{\selectlanguage{english} {}-} &
{\selectlanguage{english} 3.85} &
{\selectlanguage{english} 7.8} &
{\selectlanguage{english} 4.30} &
{\selectlanguage{english} 4.3}\\
{\selectlanguage{english} U} &
{\selectlanguage{english}\sffamily \foreignlanguage{english}{\textrm{{}-~}}} &
{\selectlanguage{english} {}-} &
{\selectlanguage{english} {}-} &
{\selectlanguage{english} 2.70} &
{\selectlanguage{english} 13.6} &
{\selectlanguage{english} {}-} &
{\selectlanguage{english}\sffamily \foreignlanguage{english}{\textrm{~-}}}\\
{\selectlanguage{english} Eu} &
{\selectlanguage{english}\sffamily \foreignlanguage{english}{\textrm{{}-~}}} &
{\selectlanguage{english} {}-} &
{\selectlanguage{english} {}-} &
{\selectlanguage{english} {}0.54*} &
{\selectlanguage{english} {}23*} &
{\selectlanguage{english} 0.59} &
{\selectlanguage{english} 4.4}\\
{\selectlanguage{english} Hf} &
{\selectlanguage{english}\sffamily \foreignlanguage{english}{\textrm{{}-~}}} &
{\selectlanguage{english} {}-} &
{\selectlanguage{english} {}-} &
{\selectlanguage{english} {}-} &
{\selectlanguage{english} {}-} &
{\selectlanguage{english} 2.70} &
{\selectlanguage{english} 4.6}\\
{\selectlanguage{english} Tb} &
{\selectlanguage{english}\sffamily \foreignlanguage{english}{\textrm{{}-~}}} &
{\selectlanguage{english} {}-} &
{\selectlanguage{english} {}-} &
{\selectlanguage{english} {}-} &
{\selectlanguage{english} {}-} &
{\selectlanguage{english} 0.30} &
{\selectlanguage{english} 11.7}\\
{\selectlanguage{english} Yb} &
{\selectlanguage{english}\sffamily \foreignlanguage{english}{\textrm{{}-~}}} &
{\selectlanguage{english} {}-} &
{\selectlanguage{english} {}-} &
{\selectlanguage{english} {}-} &
{\selectlanguage{english} {}-} &
{\selectlanguage{english} 1.03} &
{\selectlanguage{english} 9.3}\\\hline
\end{tabular}}
\end{table}

\bigskip

\begin{table}[htbp]
  \centering
  \caption{Measured elemental composition of Reference concrete with PGAA, XRF (most of them with EDXRF, marked (*) ones with WDXRF) and NAA analytical methods. }
  \label{tab:compRef}
  \resizebox *{!}{\textheight}{
\begin{tabular}{cccccccc}
      \hline
      Element  &  Nominal  &  PGAA  &  Unc. [\%]  &  XRF  &  Unc. [\%]  &  NAA  &  Unc. [\%]  \\
      \hline
        &  \multicolumn{7}{c}{ [w\%]}  \\
      \hline 
{\selectlanguage{english} H} &
{\selectlanguage{english} 0.72} &
{\selectlanguage{english} 0.60} &
{\selectlanguage{english} 1.6} &
{\selectlanguage{english} {}-} &
{\selectlanguage{english}\sffamily \foreignlanguage{english}{\textrm{{}-~}}} &
{\selectlanguage{english} {}-} &
{\selectlanguage{english}\sffamily \foreignlanguage{english}{\textrm{{}-~}}}\\
{\selectlanguage{english} Na} &
{\selectlanguage{english} 1.06} &
{\selectlanguage{english} 2.00} &
{\selectlanguage{english} 2.4} &
{\selectlanguage{english} 1.97} &
{\selectlanguage{english} 10} &
{\selectlanguage{english} 1.99} &
{\selectlanguage{english} 3.7}\\
{\selectlanguage{english} Al} &
{\selectlanguage{english} 3.70} &
{\selectlanguage{english} 5.29} &
{\selectlanguage{english} 2.2} &
{\selectlanguage{english} 6.66} &
{\selectlanguage{english} 0.5} &
{\selectlanguage{english} {}-} &
{\selectlanguage{english}\sffamily \foreignlanguage{english}{\textrm{{}-~}}}\\
{\selectlanguage{english} Si} &
{\selectlanguage{english} 32.7} &
{\selectlanguage{english}\sffamily \foreignlanguage{english}{\textrm{31.3}}} &
{\selectlanguage{english} 0.8} &
{\selectlanguage{english} 30.1} &
{\selectlanguage{english}\sffamily \foreignlanguage{english}{\textrm{0.1}}} &
{\selectlanguage{english} {}-} &
{\selectlanguage{english}\sffamily \foreignlanguage{english}{\textrm{{}-~}}}\\
{\selectlanguage{english} S} &
{\selectlanguage{english} 0.24} &
{\selectlanguage{english} 0.28} &
{\selectlanguage{english} 3.8} &
{\selectlanguage{english} {}-} &
{\selectlanguage{english}\sffamily \foreignlanguage{english}{\textrm{{}-~}}} &
{\selectlanguage{english} {}-} &
{\selectlanguage{english}\sffamily \foreignlanguage{english}{\textrm{{}-~}}}\\
{\selectlanguage{english} K} &
{\selectlanguage{english} 2.12} &
{\selectlanguage{english} 2.13} &
{\selectlanguage{english} 2.2} &
{\selectlanguage{english} 2.19} &
{\selectlanguage{english}\sffamily \foreignlanguage{english}{\textrm{0.1}}} &
{\selectlanguage{english} {}-} &
{\selectlanguage{english}\sffamily \foreignlanguage{english}{\textrm{{}-~}}}\\
{\selectlanguage{english} Ca} &
{\selectlanguage{english} 7.12} &
{\selectlanguage{english} 7.08} &
{\selectlanguage{english} 2.8} &
{\selectlanguage{english} 6.63} &
{\selectlanguage{english}\sffamily \foreignlanguage{english}{\textrm{0.1}}} &
{\selectlanguage{english} {}-} &
{\selectlanguage{english}\sffamily \foreignlanguage{english}{\textrm{{}-~}}}\\
{\selectlanguage{english} Ti} &
{\selectlanguage{english} 0.09} &
{\selectlanguage{english} 0.16} &
{\selectlanguage{english} 2.5} &
{\selectlanguage{english} 0.18} &
{\selectlanguage{english} 0.6} &
{\selectlanguage{english} {}-} &
{\selectlanguage{english} {}-}\\
{\selectlanguage{english} Mn} &
{\selectlanguage{english}\sffamily \foreignlanguage{english}{\textrm{{}-~}}} &
{\selectlanguage{english} 0.03} &
{\selectlanguage{english} 3.5} &
{\selectlanguage{english} 0.02} &
{\selectlanguage{english} 4.3} &
{\selectlanguage{english} {}-} &
{\selectlanguage{english} {}-}\\
{\selectlanguage{english} Fe} &
{\selectlanguage{english} 1.16} &
{\selectlanguage{english} 1.36} &
{\selectlanguage{english} 2.6} &
{\selectlanguage{english} 1.34} &
{\selectlanguage{english}\sffamily \foreignlanguage{english}{\textrm{0.1}}} &
{\selectlanguage{english} 1.55} &
{\selectlanguage{english} 3.8}\\
{\selectlanguage{english} O} &
{\selectlanguage{english} 50.8} &
{\selectlanguage{english} 50.2} &
{\selectlanguage{english} 4} &
{\selectlanguage{english}\sffamily \foreignlanguage{english}{\textrm{{}-~}}} &
{\selectlanguage{english}\sffamily \foreignlanguage{english}{\textrm{{}-~}}} &
{\selectlanguage{english}\sffamily \foreignlanguage{english}{\textrm{~-}}} &
{\selectlanguage{english}\sffamily \foreignlanguage{english}{\textrm{~-}}}\\
      \hline
       &  \multicolumn{7}{c}{  [mg/kg] }  \\  
      \hline 
{\selectlanguage{english} B} &
{\selectlanguage{english}\sffamily \foreignlanguage{english}{\textrm{{}-~}}} &
{\selectlanguage{english} 24.9} &
{\selectlanguage{english} 1.2} &
{\selectlanguage{english} {}-} &
{\selectlanguage{english}\sffamily \foreignlanguage{english}{\textrm{{}-~}}} &
{\selectlanguage{english} {}-} &
{\selectlanguage{english}\sffamily \foreignlanguage{english}{\textrm{{}-~}}}\\
{\selectlanguage{english} Cl} &
{\selectlanguage{english} 30} &
{\selectlanguage{english} 169} &
{\selectlanguage{english} 2.9} &
{\selectlanguage{english} {}-} &
{\selectlanguage{english}\sffamily \foreignlanguage{english}{\textrm{{}-~}}} &
{\selectlanguage{english} {}-} &
{\selectlanguage{english}\sffamily \foreignlanguage{english}{\textrm{{}-~}}}\\
{\selectlanguage{english} V} &
{\selectlanguage{english}\sffamily \foreignlanguage{english}{\textrm{{}-~}}} &
{\selectlanguage{english} 70.0} &
{\selectlanguage{english}\sffamily \foreignlanguage{english}{\textrm{7.0}}} &
{\selectlanguage{english} 55.5} &
{\selectlanguage{english} 0.4} &
{\selectlanguage{english} {}-} &
{\selectlanguage{english}\sffamily \foreignlanguage{english}{\textrm{{}-~}}}\\
{\selectlanguage{english} Sm} &
{\selectlanguage{english}\sffamily \foreignlanguage{english}{\textrm{{}-~}}} &
{\selectlanguage{english} 1.97} &
{\selectlanguage{english} 1.8} &
{\selectlanguage{english} {}-} &
{\selectlanguage{english}\sffamily \foreignlanguage{english}{\textrm{{}-~}}} &
{\selectlanguage{english} 2.38} &
{\selectlanguage{english} 4.1}\\
{\selectlanguage{english} Gd} &
{\selectlanguage{english}\sffamily \foreignlanguage{english}{\textrm{{}-~}}} &
{\selectlanguage{english} 2.20} &
{\selectlanguage{english} 6.0} &
{\selectlanguage{english} {}-} &
{\selectlanguage{english}\sffamily \foreignlanguage{english}{\textrm{{}-~}}} &
{\selectlanguage{english} {}-} &
{\selectlanguage{english}\sffamily \foreignlanguage{english}{\textrm{{}-~}}}\\
{\selectlanguage{english} Mg} &
{\selectlanguage{english} 2370} &
{\selectlanguage{english} {}-} &
{\selectlanguage{english}\sffamily \foreignlanguage{english}{\textrm{{}-~}}} &
{\selectlanguage{english} 9530} &
{\selectlanguage{english} 31} &
{\selectlanguage{english} {}-} &
{\selectlanguage{english}\sffamily \foreignlanguage{english}{\textrm{{}-~}}}\\
{\selectlanguage{english} Sc} &
{\selectlanguage{english}\sffamily \foreignlanguage{english}{\textrm{{}-~}}} &
{\selectlanguage{english} {}-} &
{\selectlanguage{english}\sffamily \foreignlanguage{english}{\textrm{~-}}} &
{\selectlanguage{english}\sffamily \foreignlanguage{english}{\textrm{7.1*}}} &
{\selectlanguage{english} 25*} &
{\selectlanguage{english} 3.97} &
{\selectlanguage{english} 3.6}\\
{\selectlanguage{english} Cr} &
{\selectlanguage{english}\sffamily \foreignlanguage{english}{\textrm{{}-~}}} &
{\selectlanguage{english} {}-} &
{\selectlanguage{english} {}-} &
{\selectlanguage{english} 44.3} &
{\selectlanguage{english} 0.7} &
{\selectlanguage{english} 65.7} &
{\selectlanguage{english} 4.2}\\
{\selectlanguage{english} Co} &
{\selectlanguage{english}\sffamily \foreignlanguage{english}{\textrm{{}-~}}} &
{\selectlanguage{english} {}-} &
{\selectlanguage{english} {}-} &
{\selectlanguage{english} {}7.0*} &
{\selectlanguage{english}\sffamily \foreignlanguage{english}{\textrm{25*}}} &
{\selectlanguage{english} 5.52} &
{\selectlanguage{english} 4.4}\\
{\selectlanguage{english} Ni} &
{\selectlanguage{english}\sffamily \foreignlanguage{english}{\textrm{{}-~}}} &
{\selectlanguage{english} {}-} &
{\selectlanguage{english} {}-} &
{\selectlanguage{english} 6.40} &
{\selectlanguage{english}\sffamily \foreignlanguage{english}{\textrm{{}-~}}} &
{\selectlanguage{english} {}-} &
{\selectlanguage{english}\sffamily \foreignlanguage{english}{\textrm{{}-~}}}\\
{\selectlanguage{english} Cu} &
{\selectlanguage{english}\sffamily \foreignlanguage{english}{\textrm{{}-~}}} &
{\selectlanguage{english} {}-} &
{\selectlanguage{english} {}-} &
{\selectlanguage{english} 22.4} &
{\selectlanguage{english} 2.1} &
{\selectlanguage{english} {}-} &
{\selectlanguage{english}\sffamily \foreignlanguage{english}{\textrm{~-}}}\\
{\selectlanguage{english} Zn} &
{\selectlanguage{english}\sffamily \foreignlanguage{english}{\textrm{{}-~}}} &
{\selectlanguage{english} {}-} &
{\selectlanguage{english} {}-} &
{\selectlanguage{english}\sffamily \foreignlanguage{english}{\textrm{87.7}}} &
{\selectlanguage{english} 0.5} &
{\selectlanguage{english}\sffamily \foreignlanguage{english}{\textrm{107}}} &
{\selectlanguage{english} 6.0}\\
{\selectlanguage{english} Ga} &
{\selectlanguage{english}\sffamily \foreignlanguage{english}{\textrm{{}-~}}} &
{\selectlanguage{english} {}-} &
{\selectlanguage{english} {}-} &
{\selectlanguage{english} 10.9} &
{\selectlanguage{english} 5.7} &
{\selectlanguage{english} {}-} &
{\selectlanguage{english}\sffamily \foreignlanguage{english}{\textrm{~-}}}\\
{\selectlanguage{english} Ge} &
{\selectlanguage{english}\sffamily \foreignlanguage{english}{\textrm{{}-~}}} &
{\selectlanguage{english} {}-} &
{\selectlanguage{english} {}-} &
{\selectlanguage{english} 3.20} &
{\selectlanguage{english} 7.1} &
{\selectlanguage{english} {}-} &
{\selectlanguage{english}\sffamily \foreignlanguage{english}{\textrm{~-}}}\\
{\selectlanguage{english} As} &
{\selectlanguage{english}\sffamily \foreignlanguage{english}{\textrm{{}-~}}} &
{\selectlanguage{english} {}-} &
{\selectlanguage{english} {}-} &
{\selectlanguage{english} {}-} &
{\selectlanguage{english}\sffamily \foreignlanguage{english}{\textrm{{}-~}}} &
{\selectlanguage{english} {}-} &
{\selectlanguage{english}\sffamily \foreignlanguage{english}{\textrm{~-}}}\\
{\selectlanguage{english} Rb} &
{\selectlanguage{english}\sffamily \foreignlanguage{english}{\textrm{{}-~}}} &
{\selectlanguage{english} {}-} &
{\selectlanguage{english} {}-} &
{\selectlanguage{english}\sffamily \foreignlanguage{english}{\textrm{76.0}}} &
{\selectlanguage{english} 1.0} &
{\selectlanguage{english}\sffamily \foreignlanguage{english}{\textrm{86.2}}} &
{\selectlanguage{english} 6.4}\\
{\selectlanguage{english} Sr} &
{\selectlanguage{english}\sffamily \foreignlanguage{english}{\textrm{{}-~}}} &
{\selectlanguage{english} {}-} &
{\selectlanguage{english} {}-} &
{\selectlanguage{english} 380} &
{\selectlanguage{english} 0.4} &
{\selectlanguage{english} {}-} &
{\selectlanguage{english}\sffamily \foreignlanguage{english}{\textrm{~-}}}\\
{\selectlanguage{english} Y} &
{\selectlanguage{english}\sffamily \foreignlanguage{english}{\textrm{{}-~}}} &
{\selectlanguage{english} {}-} &
{\selectlanguage{english} {}-} &
{\selectlanguage{english}\sffamily \foreignlanguage{english}{\textrm{12.4}}} &
{\selectlanguage{english}\sffamily \foreignlanguage{english}{\textrm{3.4}}} &
{\selectlanguage{english} {}-} &
{\selectlanguage{english}\sffamily \foreignlanguage{english}{\textrm{~-}}}\\
{\selectlanguage{english} Zr} &
{\selectlanguage{english}\sffamily \foreignlanguage{english}{\textrm{{}-~}}} &
{\selectlanguage{english} {}-} &
{\selectlanguage{english} {}-} &
{\selectlanguage{english} 114} &
{\selectlanguage{english} 0.7} &
{\selectlanguage{english} {}-} &
{\selectlanguage{english}\sffamily \foreignlanguage{english}{\textrm{~-}}}\\
{\selectlanguage{english} Nb} &
{\selectlanguage{english}\sffamily \foreignlanguage{english}{\textrm{{}-~}}} &
{\selectlanguage{english} {}-} &
{\selectlanguage{english} {}-} &
{\selectlanguage{english} 6.20} &
{\selectlanguage{english} 1.0} &
{\selectlanguage{english} {}-} &
{\selectlanguage{english}\sffamily \foreignlanguage{english}{\textrm{~-}}}\\
{\selectlanguage{english} Mo} &
{\selectlanguage{english}\sffamily \foreignlanguage{english}{\textrm{{}-~}}} &
{\selectlanguage{english} {}-} &
{\selectlanguage{english} {}-} &
{\selectlanguage{english} 2.18} &
{\selectlanguage{english} 3.9} &
{\selectlanguage{english} {}-} &
{\selectlanguage{english}\sffamily \foreignlanguage{english}{\textrm{~-}}}\\
{\selectlanguage{english} In} &
{\selectlanguage{english}\sffamily \foreignlanguage{english}{\textrm{{}-~}}} &
{\selectlanguage{english} {}-} &
{\selectlanguage{english} {}-} &
{\selectlanguage{english} 1.28} &
{\selectlanguage{english} 5.1} &
{\selectlanguage{english} {}-} &
{\selectlanguage{english}\sffamily \foreignlanguage{english}{\textrm{~-}}}\\
{\selectlanguage{english} Sn} &
{\selectlanguage{english}\sffamily \foreignlanguage{english}{\textrm{{}-~}}} &
{\selectlanguage{english} {}-} &
{\selectlanguage{english} {}-} &
{\selectlanguage{english} 2.81} &
{\selectlanguage{english} 10.8} &
{\selectlanguage{english} {}-} &
{\selectlanguage{english}\sffamily \foreignlanguage{english}{\textrm{~-}}}\\
{\selectlanguage{english} Sb} &
{\selectlanguage{english}\sffamily \foreignlanguage{english}{\textrm{{}-~}}} &
{\selectlanguage{english} {}-} &
{\selectlanguage{english} {}-} &
{\selectlanguage{english} 1.25} &
{\selectlanguage{english} 12.4} &
{\selectlanguage{english} 1.25} &
{\selectlanguage{english} 7.9}\\
{\selectlanguage{english} Cs} &
{\selectlanguage{english}\sffamily \foreignlanguage{english}{\textrm{{}-~}}} &
{\selectlanguage{english} {}-} &
{\selectlanguage{english} {}-} &
{\selectlanguage{english} 3.95} &
{\selectlanguage{english} 1.6} &
{\selectlanguage{english} 2.59} &
{\selectlanguage{english} 7.8}\\
{\selectlanguage{english} Ba} &
{\selectlanguage{english}\sffamily \foreignlanguage{english}{\textrm{{}-~}}} &
{\selectlanguage{english} {}-} &
{\selectlanguage{english} {}-} &
{\selectlanguage{english} 665} &
{\selectlanguage{english} 0.2} &
{\selectlanguage{english}\sffamily \foreignlanguage{english}{\textrm{641}}} &
{\selectlanguage{english} 4.7}\\
{\selectlanguage{english} La} &
{\selectlanguage{english}\sffamily \foreignlanguage{english}{\textrm{{}-~}}} &
{\selectlanguage{english} {}-} &
{\selectlanguage{english} {}-} &
{\selectlanguage{english} 23.7} &
{\selectlanguage{english} 1.2} &
{\selectlanguage{english} 21.44} &
{\selectlanguage{english}\sffamily \foreignlanguage{english}{\textrm{3.9}}}\\
{\selectlanguage{english} Ce} &
{\selectlanguage{english}\sffamily \foreignlanguage{english}{\textrm{{}-~}}} &
{\selectlanguage{english} {}-} &
{\selectlanguage{english} {}-} &
{\selectlanguage{english}\sffamily \foreignlanguage{english}{\textrm{37.0}}} &
{\selectlanguage{english} 21.3} &
{\selectlanguage{english} 45.75} &
{\selectlanguage{english} 4.0}\\
{\selectlanguage{english} Pr} &
{\selectlanguage{english}\sffamily \foreignlanguage{english}{\textrm{{}-~}}} &
{\selectlanguage{english} {}-} &
{\selectlanguage{english} {}-} &
{\selectlanguage{english} 5.49} &
{\selectlanguage{english} 10.0} &
{\selectlanguage{english} {}-} &
{\selectlanguage{english}\sffamily \foreignlanguage{english}{\textrm{{}-~}}}\\
{\selectlanguage{english} Nd} &
{\selectlanguage{english}\sffamily \foreignlanguage{english}{\textrm{{}-~}}} &
{\selectlanguage{english} {}-} &
{\selectlanguage{english} {}-} &
{\selectlanguage{english} 18.6} &
{\selectlanguage{english} 23.1} &
{\selectlanguage{english}\sffamily \foreignlanguage{english}{\textrm{20.32}}} &
{\selectlanguage{english} 8.0}\\
{\selectlanguage{english} W} &
{\selectlanguage{english}\sffamily \foreignlanguage{english}{\textrm{{}-~}}} &
{\selectlanguage{english} {}-} &
{\selectlanguage{english} {}-} &
{\selectlanguage{english} {}-} &
{\selectlanguage{english} {}-} &
{\selectlanguage{english}\sffamily \foreignlanguage{english}{\textrm{596}}} &
{\selectlanguage{english} 3.6}\\
{\selectlanguage{english} Pb} &
{\selectlanguage{english}\sffamily \foreignlanguage{english}{\textrm{{}-~}}} &
{\selectlanguage{english} {}-} &
{\selectlanguage{english} {}-} &
{\selectlanguage{english}\sffamily \foreignlanguage{english}{\textrm{19.4}}} &
{\selectlanguage{english} 1.9} &
{\selectlanguage{english} {}-} &
{\selectlanguage{english}\sffamily \foreignlanguage{english}{\textrm{{}-~}}}\\
{\selectlanguage{english} Th} &
{\selectlanguage{english}\sffamily \foreignlanguage{english}{\textrm{{}-~}}} &
{\selectlanguage{english} {}-} &
{\selectlanguage{english} {}-} &
{\selectlanguage{english} 4.71} &
{\selectlanguage{english} 4.9} &
{\selectlanguage{english}\sffamily \foreignlanguage{english}{\textrm{7.62}}} &
{\selectlanguage{english} 4.0}\\
{\selectlanguage{english} U} &
{\selectlanguage{english}\sffamily \foreignlanguage{english}{\textrm{{}-~}}} &
{\selectlanguage{english} {}-} &
{\selectlanguage{english} {}-} &
{\selectlanguage{english} 3.39} &
{\selectlanguage{english} 4.6} &
{\selectlanguage{english} {}-} &
{\selectlanguage{english}\sffamily \foreignlanguage{english}{\textrm{{}-~}}}\\
{\selectlanguage{english} Eu} &
{\selectlanguage{english}\sffamily \foreignlanguage{english}{\textrm{{}-~}}} &
{\selectlanguage{english} {}-} &
{\selectlanguage{english} {}-} &
{\selectlanguage{english} {}0.54*} &
{\selectlanguage{english}\sffamily \foreignlanguage{english}{\textrm{23*}}} &
{\selectlanguage{english} 0.70} &
{\selectlanguage{english} 4.3}\\
{\selectlanguage{english} Hf} &
{\selectlanguage{english}\sffamily \foreignlanguage{english}{\textrm{{}-~}}} &
{\selectlanguage{english} {}-} &
{\selectlanguage{english} {}-} &
{\selectlanguage{english} {}-} &
{\selectlanguage{english} {}-} &
{\selectlanguage{english} 2.98} &
{\selectlanguage{english} 4.5}\\
{\selectlanguage{english} Tb} &
{\selectlanguage{english}\sffamily \foreignlanguage{english}{\textrm{{}-~}}} &
{\selectlanguage{english} {}-} &
{\selectlanguage{english} {}-} &
{\selectlanguage{english} {}-} &
{\selectlanguage{english} {}-} &
{\selectlanguage{english} 0.37} &
{\selectlanguage{english} 10.3}\\
{\selectlanguage{english} Yb} &
{\selectlanguage{english}\sffamily \foreignlanguage{english}{\textrm{{}-~}}} &
{\selectlanguage{english} {}-} &
{\selectlanguage{english} {}-} &
{\selectlanguage{english} {}-} &
{\selectlanguage{english} {}-} &
{\selectlanguage{english} {}-} &
{\selectlanguage{english}\sffamily \foreignlanguage{english}{\textrm{{}-~}}}\\
{\selectlanguage{english} P} &
{\selectlanguage{english} 454} &
{\selectlanguage{english} {}-} &
{\selectlanguage{english} {}-} &
{\selectlanguage{english} {}-} &
{\selectlanguage{english} {}-} &
{\selectlanguage{english} {}-} &
{\selectlanguage{english}\sffamily \foreignlanguage{english}{\textrm{{}-~}}}\\
{\selectlanguage{english} Ta} &
{\selectlanguage{english}\sffamily \foreignlanguage{english}{\textrm{~-}}} &
{\selectlanguage{english} {}-} &
{\selectlanguage{english} {}-} &
{\selectlanguage{english} {}-} &
{\selectlanguage{english} {}-} &
{\selectlanguage{english} 0.62} &
{\selectlanguage{english} 7.4}\\\hline
\end{tabular}}
\end{table}

\bigskip

\begin{table}[htbp]
  \centering
  \caption{Measured elemental composition of Skanska concrete with PGAA, XRF (most of them with EDXRF, marked (*) ones with WDXRF) and NAA analytical methods. }
  \label{tab:compSka}
  \resizebox *{!}{\textheight}{
\begin{tabular}{cccccccc}
      \hline
      Element  &  PGAA  &  Unc. [\%]  &  XRF  &  Unc. [\%]  &  NAA  &  Unc. [\%]  \\
      \hline
        &  \multicolumn{6}{c}{ [w\%]}  \\
      \hline 
{\selectlanguage{english} H} &
{\selectlanguage{english} 0.35} &
{\selectlanguage{english} 2.1} &
{\selectlanguage{english}\sffamily \foreignlanguage{english}{\textrm{{}-~}}} &
{\selectlanguage{english}\sffamily \foreignlanguage{english}{\textrm{~-}}} &
{\selectlanguage{english} {}-} &
{\selectlanguage{english}\sffamily \foreignlanguage{english}{\textrm{{}-~}}}\\
{\selectlanguage{english} C} &
{\selectlanguage{english} 0.55} &
{\selectlanguage{english} 57.1} &
{\selectlanguage{english}\sffamily \foreignlanguage{english}{\textrm{~-}}} &
{\selectlanguage{english}\sffamily \foreignlanguage{english}{\textrm{~-}}} &
{\selectlanguage{english} {}-} &
{\selectlanguage{english}\sffamily \foreignlanguage{english}{\textrm{{}-~}}}\\
{\selectlanguage{english} Na} &
{\selectlanguage{english} 0.55} &
{\selectlanguage{english} 2.7} &
{\selectlanguage{english} 0.5} &
{\selectlanguage{english} 46} &
{\selectlanguage{english} 0.56} &
{\selectlanguage{english} 3.8}\\
{\selectlanguage{english} Al} &
{\selectlanguage{english} 2.59} &
{\selectlanguage{english} 2.6} &
{\selectlanguage{english}\sffamily \foreignlanguage{english}{\textrm{4.9}}} &
{\selectlanguage{english} 15.7} &
{\selectlanguage{english} {}-} &
{\selectlanguage{english}\sffamily \foreignlanguage{english}{\textrm{{}-~}}}\\
{\selectlanguage{english} Si} &
{\selectlanguage{english} 34.1} &
{\selectlanguage{english} 1.3} &
{\selectlanguage{english}\sffamily \foreignlanguage{english}{\textrm{29.9}}} &
{\selectlanguage{english} 4.3} &
{\selectlanguage{english} {}-} &
{\selectlanguage{english}\sffamily \foreignlanguage{english}{\textrm{{}-~}}}\\
{\selectlanguage{english} S} &
{\selectlanguage{english} 0.31} &
{\selectlanguage{english} 3.2} &
{\selectlanguage{english} 0.258} &
{\selectlanguage{english} 2.5} &
{\selectlanguage{english} {}-} &
{\selectlanguage{english}\sffamily \foreignlanguage{english}{\textrm{{}-~}}}\\
{\selectlanguage{english} K} &
{\selectlanguage{english} 1.28} &
{\selectlanguage{english}\sffamily \foreignlanguage{english}{\textrm{2.6}}} &
{\selectlanguage{english} 1.48} &
{\selectlanguage{english} 0.6} &
{\selectlanguage{english} {}-} &
{\selectlanguage{english}\sffamily \foreignlanguage{english}{\textrm{{}-~}}}\\
{\selectlanguage{english} Ca} &
{\selectlanguage{english} 8.36} &
{\selectlanguage{english} 3.0} &
{\selectlanguage{english} 9.21} &
{\selectlanguage{english} 1.1} &
{\selectlanguage{english} {}-} &
{\selectlanguage{english}\sffamily \foreignlanguage{english}{\textrm{{}-~}}}\\
{\selectlanguage{english} Ti} &
{\selectlanguage{english} 0.16} &
{\selectlanguage{english} 3.1} &
{\selectlanguage{english} 0.187} &
{\selectlanguage{english} 2.7} &
{\selectlanguage{english} {}-} &
{\selectlanguage{english}\sffamily \foreignlanguage{english}{\textrm{{}-~}}}\\
{\selectlanguage{english} Mn} &
{\selectlanguage{english} 0.05} &
{\selectlanguage{english} 3.2} &
{\selectlanguage{english} 0.045} &
{\selectlanguage{english} 2.2} &
{\selectlanguage{english} {}-} &
{\selectlanguage{english}\sffamily \foreignlanguage{english}{\textrm{{}-~}}}\\
{\selectlanguage{english} Fe} &
{\selectlanguage{english} 1.61} &
{\selectlanguage{english} 3.2} &
{\selectlanguage{english}\sffamily \foreignlanguage{english}{\textrm{1.64}}} &
{\selectlanguage{english}\sffamily \foreignlanguage{english}{\textrm{2.4}}} &
{\selectlanguage{english} 2.10} &
{\selectlanguage{english} 3.7}\\
{\selectlanguage{english} O} &
{\selectlanguage{english} 50.46} &
{\selectlanguage{english} 3} &
{\selectlanguage{english}\sffamily \foreignlanguage{english}{\textrm{{}-~}}} &
{\selectlanguage{english}\sffamily \foreignlanguage{english}{\textrm{{}-~}}} &
{\selectlanguage{english}\sffamily \foreignlanguage{english}{\textrm{~-}}} &
{\selectlanguage{english}\sffamily \foreignlanguage{english}{\textrm{{}-~}}}\\
      \hline
       &  \multicolumn{6}{c}{  [mg/kg]  }  \\  
      \hline 
{\selectlanguage{english} B} &
{\selectlanguage{english} 25.2} &
{\selectlanguage{english} 1.3} &
{\selectlanguage{english}\sffamily \foreignlanguage{english}{\textrm{{}-~}}} &
{\selectlanguage{english}\sffamily \foreignlanguage{english}{\textrm{{}-~}}} &
{\selectlanguage{english} {}-} &
{\selectlanguage{english}\sffamily \foreignlanguage{english}{\textrm{{}-~}}}\\
{\selectlanguage{english} Cl} &
{\selectlanguage{english} 151} &
{\selectlanguage{english} 1.8} &
{\selectlanguage{english} 140} &
{\selectlanguage{english} 7.1} &
{\selectlanguage{english} {}-} &
{\selectlanguage{english}\sffamily \foreignlanguage{english}{\textrm{{}-~}}}\\
{\selectlanguage{english} V} &
{\selectlanguage{english} {}-} &
{\selectlanguage{english}\sffamily \foreignlanguage{english}{\textrm{{}-~}}} &
{\selectlanguage{english} 64} &
{\selectlanguage{english} 11.7} &
{\selectlanguage{english} {}-} &
{\selectlanguage{english}\sffamily \foreignlanguage{english}{\textrm{{}-~}}}\\
{\selectlanguage{english} Sm} &
{\selectlanguage{english} 1.7} &
{\selectlanguage{english} 1.9} &
{\selectlanguage{english}\sffamily \foreignlanguage{english}{\textrm{{}-~}}} &
{\selectlanguage{english}\sffamily \foreignlanguage{english}{\textrm{{}-~}}} &
{\selectlanguage{english} 2.25} &
{\selectlanguage{english} 4.0}\\
{\selectlanguage{english} Gd} &
{\selectlanguage{english} 2.2} &
{\selectlanguage{english} 6.0} &
{\selectlanguage{english}\sffamily \foreignlanguage{english}{\textrm{{}-~}}} &
{\selectlanguage{english}\sffamily \foreignlanguage{english}{\textrm{{}-~}}} &
{\selectlanguage{english} {}-} &
{\selectlanguage{english}\sffamily \foreignlanguage{english}{\textrm{{}-~}}}\\
{\selectlanguage{english} Mg} &
{\selectlanguage{english} {}-} &
{\selectlanguage{english}\sffamily \foreignlanguage{english}{\textrm{~-}}} &
{\selectlanguage{english}\sffamily \foreignlanguage{english}{\textrm{6360}}} &
{\selectlanguage{english} 15.7} &
{\selectlanguage{english} {}-} &
{\selectlanguage{english}\sffamily \foreignlanguage{english}{\textrm{{}-~}}}\\
{\selectlanguage{english} Sc} &
{\selectlanguage{english} {}-} &
{\selectlanguage{english}\sffamily \foreignlanguage{english}{\textrm{~-}}} &
{\selectlanguage{english}\sffamily \foreignlanguage{english}{\textrm{8.3*}}} &
{\selectlanguage{english}\sffamily \foreignlanguage{english}{\textrm{22*}}} &
{\selectlanguage{english} 4.75} &
{\selectlanguage{english} 3.6}\\
{\selectlanguage{english} Cr} &
{\selectlanguage{english} {}-} &
{\selectlanguage{english} {}-} &
{\selectlanguage{english} 63.6} &
{\selectlanguage{english} 3.5} &
{\selectlanguage{english}\sffamily \foreignlanguage{english}{\textrm{110}}} &
{\selectlanguage{english} 3.9}\\
{\selectlanguage{english} Co} &
{\selectlanguage{english} {}-} &
{\selectlanguage{english} {}-} &
{\selectlanguage{english} {}6.8*} &
{\selectlanguage{english}\sffamily \foreignlanguage{english}{\textrm{22*}}} &
{\selectlanguage{english} 7.69} &
{\selectlanguage{english} 4.1}\\
{\selectlanguage{english} Ni} &
{\selectlanguage{english} {}-} &
{\selectlanguage{english} {}-} &
{\selectlanguage{english} 9.3} &
{\selectlanguage{english} 6.3} &
{\selectlanguage{english} {}-} &
{\selectlanguage{english}\sffamily \foreignlanguage{english}{\textrm{{}-~}}}\\
{\selectlanguage{english} Cu} &
{\selectlanguage{english} {}-} &
{\selectlanguage{english} {}-} &
{\selectlanguage{english} 17.3} &
{\selectlanguage{english} 4.9} &
{\selectlanguage{english} {}-} &
{\selectlanguage{english}\sffamily \foreignlanguage{english}{\textrm{{}-~}}}\\
{\selectlanguage{english} Zn} &
{\selectlanguage{english} {}-} &
{\selectlanguage{english} {}-} &
{\selectlanguage{english} 53.4} &
{\selectlanguage{english} 3.4} &
{\selectlanguage{english}\sffamily \foreignlanguage{english}{\textrm{65.7}}} &
{\selectlanguage{english} 8.0}\\
{\selectlanguage{english} Ga} &
{\selectlanguage{english} {}-} &
{\selectlanguage{english} {}-} &
{\selectlanguage{english} 4.0} &
{\selectlanguage{english} 6.5} &
{\selectlanguage{english} {}-} &
{\selectlanguage{english}\sffamily \foreignlanguage{english}{\textrm{{}-~}}}\\
{\selectlanguage{english} Ge} &
{\selectlanguage{english} {}-} &
{\selectlanguage{english} {}-} &
{\selectlanguage{english} 1.7} &
{\selectlanguage{english} 13.8} &
{\selectlanguage{english} {}-} &
{\selectlanguage{english}\sffamily \foreignlanguage{english}{\textrm{{}-~}}}\\
{\selectlanguage{english} As} &
{\selectlanguage{english} {}-} &
{\selectlanguage{english} {}-} &
{\selectlanguage{english} 2.0} &
{\selectlanguage{english} 4.9} &
{\selectlanguage{english} {}-} &
{\selectlanguage{english}\sffamily \foreignlanguage{english}{\textrm{{}-~}}}\\
{\selectlanguage{english} Rb} &
{\selectlanguage{english} {}-} &
{\selectlanguage{english} {}-} &
{\selectlanguage{english} 54.6} &
{\selectlanguage{english} 1.1} &
{\selectlanguage{english} 65.9} &
{\selectlanguage{english} 7.8}\\
{\selectlanguage{english} Sr} &
{\selectlanguage{english} {}-} &
{\selectlanguage{english} {}-} &
{\selectlanguage{english}\sffamily \foreignlanguage{english}{\textrm{171}}} &
{\selectlanguage{english} 0.7} &
{\selectlanguage{english} {}-} &
{\selectlanguage{english}\sffamily \foreignlanguage{english}{\textrm{{}-~}}}\\
{\selectlanguage{english} Y} &
{\selectlanguage{english} {}-} &
{\selectlanguage{english} {}-} &
{\selectlanguage{english} 15.1} &
{\selectlanguage{english} 2.7} &
{\selectlanguage{english} {}-} &
{\selectlanguage{english}\sffamily \foreignlanguage{english}{\textrm{{}-~}}}\\
{\selectlanguage{english} Zr} &
{\selectlanguage{english} {}-} &
{\selectlanguage{english} {}-} &
{\selectlanguage{english} 145} &
{\selectlanguage{english} 0.3} &
{\selectlanguage{english} {}-} &
{\selectlanguage{english}\sffamily \foreignlanguage{english}{\textrm{{}-~}}}\\
{\selectlanguage{english} Nb} &
{\selectlanguage{english} {}-} &
{\selectlanguage{english} {}-} &
{\selectlanguage{english} 4.9} &
{\selectlanguage{english}\sffamily \foreignlanguage{english}{\textrm{2.3}}} &
{\selectlanguage{english} {}-} &
{\selectlanguage{english}\sffamily \foreignlanguage{english}{\textrm{{}-~}}}\\
{\selectlanguage{english} Mo} &
{\selectlanguage{english} {}-} &
{\selectlanguage{english} {}-} &
{\selectlanguage{english} 3.0} &
{\selectlanguage{english} 5.6} &
{\selectlanguage{english} {}-} &
{\selectlanguage{english}\sffamily \foreignlanguage{english}{\textrm{{}-~}}}\\
{\selectlanguage{english} In} &
{\selectlanguage{english} {}-} &
{\selectlanguage{english} {}-} &
{\selectlanguage{english} 1.1} &
{\selectlanguage{english} 9.4} &
{\selectlanguage{english} {}-} &
{\selectlanguage{english}\sffamily \foreignlanguage{english}{\textrm{{}-~}}}\\
{\selectlanguage{english} Sn} &
{\selectlanguage{english} {}-} &
{\selectlanguage{english} {}-} &
{\selectlanguage{english} 1.5} &
{\selectlanguage{english} 18.0} &
{\selectlanguage{english} {}-} &
{\selectlanguage{english}\sffamily \foreignlanguage{english}{\textrm{{}-~}}}\\
{\selectlanguage{english} Sb} &
{\selectlanguage{english} {}-} &
{\selectlanguage{english} {}-} &
{\selectlanguage{english} 1.3} &
{\selectlanguage{english} 15.3} &
{\selectlanguage{english} 1.33} &
{\selectlanguage{english} 11.3}\\
{\selectlanguage{english} Cs} &
{\selectlanguage{english} {}-} &
{\selectlanguage{english} {}-} &
{\selectlanguage{english} 1.9} &
{\selectlanguage{english} 11.3} &
{\selectlanguage{english} 1.10} &
{\selectlanguage{english} 10.9}\\
{\selectlanguage{english} Ba} &
{\selectlanguage{english} {}-} &
{\selectlanguage{english} {}-} &
{\selectlanguage{english}\sffamily \foreignlanguage{english}{\textrm{387}}} &
{\selectlanguage{english} 0.4} &
{\selectlanguage{english}\sffamily \foreignlanguage{english}{\textrm{405}}} &
{\selectlanguage{english} 5.9}\\
{\selectlanguage{english} La} &
{\selectlanguage{english} {}-} &
{\selectlanguage{english} {}-} &
{\selectlanguage{english} 16.1} &
{\selectlanguage{english} 2.7} &
{\selectlanguage{english} 14.6} &
{\selectlanguage{english} 3.9}\\
{\selectlanguage{english} Ce} &
{\selectlanguage{english} {}-} &
{\selectlanguage{english} {}-} &
{\selectlanguage{english} 28.5} &
{\selectlanguage{english} 2.1} &
{\selectlanguage{english} 38.9} &
{\selectlanguage{english} 4.2}\\
{\selectlanguage{english} Pr} &
{\selectlanguage{english} {}-} &
{\selectlanguage{english} {}-} &
{\selectlanguage{english} 4.9} &
{\selectlanguage{english}\sffamily \foreignlanguage{english}{\textrm{11.3}}} &
{\selectlanguage{english} {}-} &
{\selectlanguage{english}\sffamily \foreignlanguage{english}{\textrm{{}-~}}}\\
{\selectlanguage{english} Nd} &
{\selectlanguage{english} {}-} &
{\selectlanguage{english} {}-} &
{\selectlanguage{english} 16.7} &
{\selectlanguage{english} 5.6} &
{\selectlanguage{english} 17.29} &
{\selectlanguage{english} 9.1}\\
{\selectlanguage{english} W} &
{\selectlanguage{english} {}-} &
{\selectlanguage{english} {}-} &
{\selectlanguage{english} {}-} &
{\selectlanguage{english}\sffamily \foreignlanguage{english}{\textrm{~-}}} &
{\selectlanguage{english}\sffamily \foreignlanguage{english}{\textrm{175}}} &
{\selectlanguage{english}\sffamily \foreignlanguage{english}{\textrm{3.8}}}\\
{\selectlanguage{english} Pb} &
{\selectlanguage{english} {}-} &
{\selectlanguage{english} {}-} &
{\selectlanguage{english} 12.6} &
{\selectlanguage{english} 4.7} &
{\selectlanguage{english} {}-} &
{\selectlanguage{english}\sffamily \foreignlanguage{english}{\textrm{{}-~}}}\\
{\selectlanguage{english} Th} &
{\selectlanguage{english} {}-} &
{\selectlanguage{english} {}-} &
{\selectlanguage{english} 3.0} &
{\selectlanguage{english} 7.7} &
{\selectlanguage{english} 6.10} &
{\selectlanguage{english} 4.1}\\
{\selectlanguage{english} U} &
{\selectlanguage{english} {}-} &
{\selectlanguage{english} {}-} &
{\selectlanguage{english}\sffamily \foreignlanguage{english}{\textrm{{}-~}}} &
{\selectlanguage{english}\sffamily \foreignlanguage{english}{\textrm{~-}}} &
{\selectlanguage{english} {}-} &
{\selectlanguage{english}\sffamily \foreignlanguage{english}{\textrm{{}-~}}}\\
{\selectlanguage{english} Eu} &
{\selectlanguage{english} {}-} &
{\selectlanguage{english} {}-} &
{\selectlanguage{english}\sffamily \foreignlanguage{english}{\textrm{0.56*}}} &
{\selectlanguage{english}\sffamily \foreignlanguage{english}{\textrm{23*}}} &
{\selectlanguage{english} 0.62} &
{\selectlanguage{english} 4.5}\\
{\selectlanguage{english} Hf} &
{\selectlanguage{english} {}-} &
{\selectlanguage{english} {}-} &
{\selectlanguage{english}\sffamily \foreignlanguage{english}{\textrm{{}-~}}} &
{\selectlanguage{english} {}-} &
{\selectlanguage{english} 4.70} &
{\selectlanguage{english} 5.9}\\
{\selectlanguage{english}\sffamily \foreignlanguage{english}{\textrm{Tb}}} &
{\selectlanguage{english} {}-} &
{\selectlanguage{english} {}-} &
{\selectlanguage{english}\sffamily \foreignlanguage{english}{\textrm{{}-~}}} &
{\selectlanguage{english} {}-} &
{\selectlanguage{english} 0.43} &
{\selectlanguage{english} 9.4}\\
{\selectlanguage{english} Yb} &
{\selectlanguage{english} {}-} &
{\selectlanguage{english} {}-} &
{\selectlanguage{english}\sffamily \foreignlanguage{english}{\textrm{{}-~}}} &
{\selectlanguage{english} {}-} &
{\selectlanguage{english} 1.69} &
{\selectlanguage{english} 5.4}\\
{\selectlanguage{english} P} &
{\selectlanguage{english} {}-} &
{\selectlanguage{english} {}-} &
{\selectlanguage{english}\sffamily \foreignlanguage{english}{\textrm{{}-~}}} &
{\selectlanguage{english} {}-} &
{\selectlanguage{english} {}-} &
{\selectlanguage{english}\sffamily \foreignlanguage{english}{\textrm{{}-~}}}\\
{\selectlanguage{english} Ta} &
{\selectlanguage{english} {}-} &
{\selectlanguage{english} {}-} &
{\selectlanguage{english}\sffamily \foreignlanguage{english}{\textrm{{}-~}}} &
{\selectlanguage{english} {}-} &
{\selectlanguage{english} 0.40} &
{\selectlanguage{english} 10}\\\hline
\end{tabular}}
\end{table}

\FloatBarrier


\end{document}